\def\paperversion{tr}
\def\anonymoussubmission{1} 
\newif\ifsinglecolumn\singlecolumnfalse
\newif\ifwidemargins\widemarginsfalse
\newif\ifwarning\warningfalse
\newif\ifshowcomments\showcommentsfalse
\newif\ifblinded\blindedfalse
\newif\ifshowlinenums\showlinenumsfalse
\newif\ifreport\reportfalse
\newif\ifcopyrightspace\copyrightspacefalse
\newif\ifacknowledgments\acknowledgmentsfalse
\newif\ifshowpagenumbers\showpagenumberstrue
\newif\iffinalformat\finalformatfalse
\newif\ifweb\webfalse

\ifx\anonymoussubmission\undefined
\else
  \if\anonymoussubmission 1
    \blindedtrue
  \else
    \blindedfalse
  \fi
\fi
\IfFileExists{.blinded}{\blindedtrue}

\ifx\paperversion\xxxxundefined
\PackageError{paperversions}{*** No valid document version was specified.
Macro paperversions must be defined as one of (markup, draft,
local, submission, final, web, tr, trdraft)}
\fi

\def\xxversion{\csname xx\paperversion\endcsname}
\newif\ifsawversion\sawversionfalse

\ifcase\xxversion\relax
    \widemarginstrue
    \singlecolumntrue
    \warningtrue
    \showlinenumstrue
    \sawversiontrue
    \blindedfalse
\or 
    \warningtrue
    \showlinenumsfalse
    \showcommentstrue
    \sawversiontrue
\or 
    \warningtrue
    \showcommentsfalse
    \blindedfalse
    \sawversiontrue
    \acknowledgmentstrue
    \showlinenumstrue
\or 
    \sawversiontrue
    \showlinenumstrue
\or 
    \blindedfalse
    \sawversiontrue
    \copyrightspacetrue
    \acknowledgmentstrue
    \showpagenumbersfalse
    \finalformattrue
\or 
    \singlecolumntrue
    \blindedfalse
    \sawversiontrue
    \reporttrue
    \acknowledgmentstrue
    \webtrue
\or 
    \blindedfalse
    \singlecolumntrue
    \showcommentstrue
    \sawversiontrue
    \reporttrue
    \showlinenumstrue
    \warningtrue
\or 
    \blindedfalse
    \showcommentstrue
    \copyrightspacetrue
    \acknowledgmentstrue
    \sawversiontrue
    \finalformattrue
    \showlinenumstrue
    \warningtrue
\or 
    \blindedfalse
    \sawversiontrue
    \copyrightspacetrue
    \acknowledgmentstrue
    \finalformattrue
    \webtrue
\or 
    \singlecolumntrue
    \blindedtrue
    \acknowledgmentsfalse
    \sawversiontrue
    \reporttrue
    \webtrue
\fi

\ifsawversion
\else
\fi

\let\xxversion=\undefined

\newif\iftableofcontents 
\newif\iftableofcontentsparagraph 
\ifreport
    \documentclass[sigplan,10pt,reprint,onecolumn]{acmart}

    \acmConference[]{Technical Report}{15 June 2018}{Cornell University}
    \acmYear{2018}
    \acmISBN{} 
    \acmDOI{} 
    \startPage{1}
    
    \setcopyright{none}
    \tableofcontentsfalse
    \tableofcontentsparagraphtrue
\else
    \tableofcontentsfalse
    \tableofcontentsparagraphtrue
    \iffinalformat
        \ifweb
            \documentclass[sigplan,10pt,reprint,onecolumn]{acmart}

            \acmConference[CCS'18]{ACM Conference on Computer and Communications Security}{October 15--19, 2018}{Toronto, Canada}
            \acmYear{2018}
            \acmISBN{} 
            \acmDOI{} 
            \startPage{1}
    
            \setcopyright{none}
            
        \else
            \documentclass[sigplan,10pt]{acmart}

            \acmConference[CCS'18]{ACM Conference on Computer and Communications Security}{October 15--19, 2018}{Toronto, Canada}
            \acmYear{2018}
            \acmISBN{} 
            \acmDOI{} 
            \startPage{1}
    
            \setcopyright{none}
        \fi
    \else
        \documentclass[sigplan,10pt,review,anonymous]{acmart}

        \acmConference[CCS'18]{ACM Conference on Computer and Communications Security}{October 15--19, 2018}{Toronto, Canada}
        \acmYear{2018}
        \acmISBN{} 
        \acmDOI{} 
        \startPage{1}
    
        \setcopyright{none}
    \fi
\fi

\usepackage{booktabs}   
\usepackage{subcaption} 

\usepackage{graphicx, color, eso-pic}
\usepackage{amsmath, amssymb, stmaryrd, utf8math}
\usepackage{tikz,pgfplots,setspace}
\usetikzlibrary{calc}
\usetikzlibrary{arrows}
\usetikzlibrary{math}

\usepackage{array, multirow}

\usepackage{hyperref}
\hypersetup{
	colorlinks,
	urlcolor=black,
	citecolor=black,
	filecolor=black,
	linkcolor=black
}

\ifshowcomments
    \usepackage[checkNoComments]{aplcomments}
\else
    \usepackage[disabled]{aplcomments}
\fi

\usepackage{listings}
\renewcommand{\floatpagefraction}{0.75}

\ifreport
\title[A Web of Blocks]{A Web of Blocks}         
\subtitle{Technical Report}             
\else
\title[Charlotte]{Charlotte: the~Decentralized Web~of~Blockchains}         
\fi

\author{Isaac Sheff}
\orcid{0000-0002-7822-1503}             
\affiliation{
  \position{PhD Candidate}
  \department{Computer Science}              
  \institution{Cornell University}            
  \city{Ithaca}
  \state{New York}
  \country{USA}                    
}
\email{isheff@cs.cornell.edu}          

\author{Xinwen Wang}
\affiliation{
  \position{PhD Student}
  \department{Computer Science}              
  \institution{Cornell University}            
  \city{Ithaca}
  \state{New York}
  \country{USA}                    
}
\email{xw467@cornell.edu}          

\author{Andrew C. Myers}
\affiliation{
  \position{Professor}
  \department{Computer Science}              
  \institution{Cornell University}            
  \city{Ithaca}
  \state{New York}
  \country{USA}                    
}
\email{andru@cs.cornell.edu}          

\author{Robbert van Renesse}
\orcid{0000-0003-3598-0283}             
\affiliation{
  \position{Research Professor}
  \department{Computer Science}              
  \institution{Cornell University}            
  \city{Ithaca}
  \state{New York}
  \country{USA}                    
}
\email{rvr@cs.cornell.edu}          

\ifwarning
\AtBeginDocument{
 \AddToShipoutPicture*{\put(240,730){\it\color{red}{\fbox{\large Draft---please do not distribute}}}}
}
\fi

\newcommand{\p}[1]{{\ensuremath{\left({{#1}}\right)}}}
\newcommand{\cb}[1]{{\left\{{{#1}}\right\}}}

\newcommand{\join}[0]{{\ensuremath{\sqcup}}}
\newcommand{\meet}[0]{{\ensuremath{\sqcap}}}

\newcommand{\blockweb}[0]{blockweb}

\usepackage{mfirstuc}
\usepackage{xspace}
\newcommand{\blockchainName}[1]{{\capitalisewords{{#1}}}}
\newcommand{\bitcoin}{{\blockchainName{bitcoin}}\xspace} 
\newcommand{\ethereum}[0]{{\blockchainName{ethereum}}\xspace} 

\newcommand{\principal}[1]{{\textsc{\capitalisewords{{#1}}}}\xspace}
\newcommand{\principalA}[0]{\principal{Alice}}
\newcommand{\principalB}[0]{\principal{Bob}}
\newcommand{\principalC}[0]{\principal{Carol}}
\newcommand{\principalD}[0]{\principal{Dave}}

\newcommand{\commodity}[1]{{\textsc{\lowercase{{#1}}}}\xspace}
\newcommand{\art}[0]{\commodity{Art}}
\newcommand{\commodityA}[0]{\art}
\newcommand{\bricks}[0]{\commodity{Bricks}}
\newcommand{\commodityB}[0]{\bricks}
\newcommand{\coffee}[0]{\commodity{Coffee}}

\newcommand{\diamonds}[0]{\commodity{Diamonds}}

\newcommand{\bitcoinendblock}[0]{{200,000}}
\newcommand{\bitcoinnumtransactions}[0]{{6,953,512}}
\newcommand{\bitcoinlengthlongest}[0]{{110,787}}
\newcommand{\bitcoinnumblocks}[0]{{\bitcoinendblock}}
\newcommand{\bitcointotaltime}[0]{{3.72 years}}

\newcommand{\bitcoincharlottetotaltime}[0]{{21.63 days}}
\newcommand{\bitcoincharlottepercenttime}[0]{{1.59{\%}}}
\newcommand{\bitcoinlengthlongestweighted}[0]{{244,163}}
\newcommand{\bitcoincharlotteweightedtotaltime}[0]{{47.68 days}}
\newcommand{\bitcoincharlotteweightedpercenttime}[0]{{3.51{\%}}}
\newcommand{\bitcoincharlotteweightedexperimenttime}[0]{{44.32 days}}
\newcommand{\bitcoincharlotteweightedexperimenttimepercent}[0]{{3.29{\%}}}

\newcommand{\bitcoinTwoTXTotal}[0]{{24,129,215}}
\newcommand{\bitcoinTwoTXTotalTime}[0]{{12.91 years}}
\newcommand{\bitcoinTwoTXExperimentTime}[0]{{12.01 years}}

\begin{document}
\begin{abstract}
  Blockchains offer a useful
 abstraction: a trustworthy, decentralized log of totally ordered
 transactions.
Traditional blockchains have problems with scalability and
 efficiency, preventing their use for many applications.
These limitations arise from the requirement that
 all participants agree on the total ordering of transactions.
To address this fundamental shortcoming, we introduce
 \textit{Charlotte}, a system for maintaining decentralized,
 authenticated data structures, including transaction logs.
Each data structure---indeed, each block---specifies its own
 availability and integrity properties, allowing Charlotte
 applications to retain the full benefits of permissioned or
 permissionless blockchains.
In Charlotte, a block can be atomically appended to multiple logs,
 allowing applications to be interoperable when they want to, without
 inefficiently \textit{forcing} all applications to share one big log.
We call this open graph of interconnected blocks a \blockweb.
We allow new kinds of {\blockweb} applications that operate beyond
 traditional chains.
We demonstrate the viability of Charlotte applications with
 proof-of-concept servers running interoperable blockchains. 
Using performance data from our prototype, we estimate that when
 compared with traditional blockchains, Charlotte offers multiple
 orders of magnitude improvement in speed and energy efficiency.

\end{abstract}
\maketitle
\iftableofcontents
  \begin{spacing}{0.91}
  \tableofcontents
  \end{spacing}
  \newpage
\fi
\iffinalformat



\fi

\newif\ifextensions
\extensionsfalse

\section{Introduction}
\label{sec:intro}
Blockchains are distributed, append-only logs used to lend
 availability, accountability, and consistency to everything from
 marketplaces~\cite{openbazaar}
 and supply chains~\cite{IBMSupplyChain}
 to health records~\cite{Marr2017}
 and governance~\cite{Wells2016}.
Despite their popularity, blockchain-based applications suffer from a
 fundamental lack of
 scalability~\cite{ScalingDecentralizedBlockchains}.
Interacting applications must be on the same
 chain, with all their operations both stored and totally
 ordered by a global mechanism.
Fully serialized blockchains are not truly scalable: each consensus
 mechanism has some maximum speed, regardless of the number of
 participating machines~\cite{ScalingDecentralizedBlockchains,
 LiuLKA2016,Miller2016}.
\ICS{
The following example is nice, but has been omitted for space:
The Ethereum Virtual Machine, for instance, can be seen as a
 single-threaded processor on which every smart contract in existence
 must share time~\cite{ethereum}.
}
As a result, popular chains like \bitcoin~\cite{bitcoin} and
 {\ethereum} are overburdened~\cite{Lee2017,Hertig2017}.

We present \textit{Charlotte}, a framework for applications built on
 the \textit{\blockweb}, a novel generalization of blockchains.
The {\blockweb} is an authenticated directed acyclic
 graph~\cite{Martel2001} of all Charlotte blocks.
  \ICS{ Martel2001 is the oldest reference I can find for a Merkle
    DAG, although it's really just the unpublished pre-print of
    Martel2004, and even then it's talking about different
    applications, and years after the earliest Merkle tree stuff. I'm
    not sure what to cite here.}
Whereas blockchains enforce a total ordering on all data, a
 {\blockweb} requires ordering only when one block references another.
Applications can thus create an ordering in the blocks they use, but
 blocks are by default unordered.
This is a natural extension of the database community's decades old
 ``least ordering'' ideal~\cite{Bernstein}.

Reduced serialization requirements make Charlotte fundamentally more
 efficient.
For example, a transaction moving money between bank accounts
 need only be serialized relative to other transactions on those
 accounts.
{\bitcoin}'s global serialization mechanism takes about an
 hour~\cite{ScalingDecentralizedBlockchains}, but with
 Charlotte, transactions can be fast enough to buy a coffee, without
 resorting to off-chain transactions settled later.
Even when operating largely in parallel, we discuss how applications
 can preserve the serializability properties of traditional
 blockchains while executing multi-chain transactions, using recursive
 attestations and meets~\p{\autoref{sec:consistency}}.

Charlotte allows applications to reference each other's blocks, and
 even to share blocks.
For example, our timestamping
 application~\p{\autoref{sec:timestamping}} reads existing blocks and
 references them with timestamp blocks, demonstrating that each
 referenced block existed before a given time.
Blocks from any other application can be sent to the timestamping
 servers, without slowing down the application.

Beyond excessive serialization, another problem is that
 existing blockchains treat all blocks the same way.
They have one mechanism for selecting and storing all blocks.
Unfortunately, no single mechanism suits every application.
In an attempt to create consensus mechanisms trustworthy enough for
 every possible application, existing blockchains have become slow and
 inefficient~\cite{ScalingDecentralizedBlockchains,bitcoinenergy}.
With Charlotte, applications specify requirements for their own
 blocks.

In Charlotte, data structures are subsets of the blocks in the
 {\blockweb}, defined by \textit{attestations} found in each block
 reference.
As each block is an immutable datum, and references others by hash,
 blocks form an authenticated directed acyclic graph~\cite{Kundu2010}.
  \ICS{ This fact is evident if you consider that when one block
    contains the hash of another, the referencer must have been made
    after the referencee, and a cyclic graph cannot be topologically
    ordered (in time).
   Kundu certainly understands this, but never outright states it to
    the reader.
   I'm not sure how we should address it so as not to go
    into unnecessary digressions.}
However, some applications require more constraints on their data
 structures.
Thus each block reference may include signed certificates
 demonstrating that it belongs in a given data structure.
These may include proofs of retrievability, demonstrating that it was
 stored properly, as well as proofs of integrity, demonstrating that
 some application-specific set of servers believe the block belongs in
 a data structure.
For instance, a payment log includes only blocks approved
 by the payer and requires a total ordering of payments, to
 protect against double spending.
Depending on the application, its data structures
 may be \textit{permissioned}, \textit{permissionless}, or
 both~\cite{Cachin2017}.
One advantage of Charlotte is that principals (or machines) that an
 application does not trust cannot influence the application's data
 structures.

With availability and integrity attestations, Charlotte separates two
 duties which traditional blockchains conflate: storing blocks,
 and approving which blocks belong in a structure~\cite{bigchaindb}.
Charlotte servers providing storage provide proofs of availability
 without having to worry about ordering or data structures, and
 \textbf{servers providing integrity need not even read whole blocks},
 \ICS{should we bold this, now that it's in the introduction?}
 streamlining consensus.
These latter servers correspond to consensus servers in a traditional
 blockchain, also called ``orderers'' or ``miners''~\cite{bigchaindb}.
As we have named our framework after the book
 \textit{Charlotte's Web}, availability servers are called
 \textit{Wilbur}, after the character whose goal is to stay
 alive, and integrity servers are called \textit{Fern}, after the
 character who makes difficult choices~\cite{White1952}.

  \iftableofcontentsparagraph
    \paragraph{Contributions}
{\samepage
\begin{itemize}
  \item In~\autoref{sec:design}, we present Charlotte, a
         {\blockweb}, which allows any application or server to
         participate concurrently.
  \item In~\autoref{sec:consistency}, we discuss the properties of
         information in Charlotte, including requirements on how
         blocks are referenced~\p{\autoref{sec:joins}} and the
         resulting properties of blockchains within
         Charlotte~\p{\autoref{sec:chain}}.
  \item In~\autoref{sec:applications}, we explain how Charlotte's
         flexibility generalizes existing
         designs~\p{\autoref{sec:simulating}} and how new
         kinds of applications can be built efficiently in Charlotte.
  \item In~\autoref{sec:implementation}, we describe our prototype
        implementation.  
  \item In~\autoref{sec:evaluation}, we use this implementation, and
         real {\bitcoin} data to compare the efficiency of Charlotte
         to that of other blockchain systems.
\ifextensions
  \item In~\autoref{sec:extensions}, we explain how Charlotte's design
         may be extended to enforce confidentiality
         properties~\p{\autoref{sec:confidentiality}} and even to allow
         specifications for acceptable block
         deletion~\p{\autoref{sec:deletion}}.
         \ACM{Maybe save this section for future work?}
\fi
\end{itemize}
}

  \fi

\section{Supply Chain: A Running Example}
\label{sec:supply}
Suppose a variety of companies want to agree on the history of
 ownership, including trades and purchases, of the goods they use.
They might want to ensure, for example, that the provenance of an
 item, or that a limited quantity was produced.~\cite{IBMSupplyChain}
A blockchain provides immutability: each record contains a digest, and
 thus a commitment, of all the records on which it depends.
If a user possesses a record, they can detect any attempted alterations
 to earlier records.
Blockchains also provide availability: each record is traditionally
 replicated with each ``miner,'' and so it is difficult to lose
 committed records.

It is possible to track supply chains on existing blockchains, such as
 \ethereum, that are intended for use as application
 platforms~\cite{ethereum}.
However, all such platforms use a single chain to track all records.
This requires that all records be totally serialized, even when
 unnecessary.
For instance, if one blockchain tracks ownership of \coffee, and
 another tracks ownership of \diamonds, then they
 should be able to operate mostly in parallel.
A sudden flurry of \coffee trades should not normally slow down trading
\diamonds.
With a single blockchain, however, that's exactly what would happen:
 trades compete to append their transactions to the
 chain~\cite{Hertig2017}.
This unnecessary ordering may seem minor, but with millions of
 potential transactions every second, even the fastest consensus
 systems cannot keep up.\ICS{Citation needed?}

One key insight is that \textbf{the mechanism for committing a block
 need not satisfy everyone: just the those involved in the block
 itself.}
Charlotte instead allows independent operations on independent servers
 to simply ignore each other.
If {\principalA} wants to trade {\commodityA} to {\principalB} in
 exchange for {\commodityB}, then any mechanism that both
 {\principalA} and {\principalB} find agreeable is sufficient to
 commit a block representing such a trade.
For example, they might agree on some consortium of servers who can
 atomically commit the block into their supply chain records.
If another trade, featuring different commodities and different
 principals, agrees on some unrelated mechanism, there is no harm in
 those trades proceeding concurrently.

Intuitively, the supply chain system needs:
\begin{itemize}
\item A way for \principalA and \principalB to state
 their requirements, in terms of who they trust to guarantee the
 uniqueness of \commodityA{} and \commodityB{} sales, and in terms of
 where such records must be stored.
\item A way for those authorities and storage servers to sign off on
 such a transaction.
\item A way for future transactions to reference this one, carrying
 attestations of uniqueness and availability.
\end{itemize}

In a trade, each party may ask for records of the history of ownership
 of the other party's holdings.
If the ownership ancestry records provided by each party are not up to
 the other's satisfaction, they will need to get their records
 \textit{endorsed} by a sufficiently trustworthy party.
Records can increase in integrity and availability as more participants
 agree to store them, or agree to reject conflicting records.

\begin{figure}
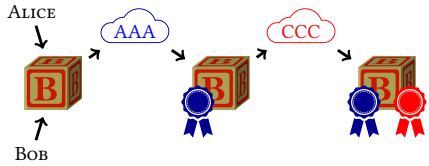

  \centering
  \include{fig-endorsement}
  \caption{\principalB buys \commodityA from \principalA, and
            the record of the sale is a block, \textit{B}.
           The first trade is tracked by the \principal{AAA}, but
            \principalB later wishes to prove ownership to
            {\principalC}, who does not trust the \principal{AAA}, and
            so they get the \principal{CCC} to \textit{endorse} the
            earlier transaction.
          }
  \label{fig:endorsement}
  \vspace{2mm}
\end{figure}

Suppose that \principalB buys some \art from \principalA, as
depicted in~\autoref{fig:endorsement}.
To ensure the authenticity of the {\art}, they atomically commit the
 change of {\art} ownership with the
 \principal{American Art Authority} (\principal{AAA}), a network of
 \art-tracking servers.
{\principalB} then has an attestation demonstrating that he owns
 the {\art}, in the eyes of the \principal{AAA}.

Suppose that {\principalB} later wishes to sell the {\art} to
 \principalC.
Naturally, the record of the new transaction should reference the
 record of the previous transaction (with {\principalA} and
 {\principalB}), to show a history of ownership.
However, {\principalC} does not trust the \principal{AAA} to maintain
 art ownership records.
She only trusts the \principal{Canadian Craft Consortium}
 (\principal{CCC}), a different network of \art-tracking servers.
Therefore, before their trade can proceed, they send the record of the
 previous transaction to the \principal{CCC}.
If the \principal{CCC} sees no conflicts, it adds it to its
 records, and replies with an attestation to that affect.
Then {\principalC} can recognize that {\principalB} owns the {\art}.

Another key insight is that \textbf{endorsement is the
 same as signing off on transactions in the first place.}
The type of atteststion the \principal{CCC} provides is no different
 from the attestation the \principal{AAA} provides.

\section{Design}
\label{sec:design}
\begin{figure}
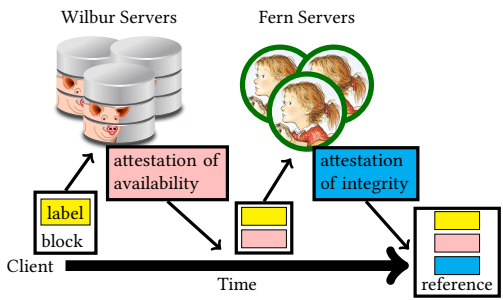

  \centering
  \include{fig-life}
  \caption{Life of a Block}
  \label{fig:life}
\end{figure}

Charlotte provides a framework for distributed applications to build
 their own data structures within the \blockweb.
Our design is motivated by the following principles:
\begin{itemize}
  \item We should not force unnecessary ordering: applications
         that do not interact should run concurrently.
  \item Storing blocks and choosing which ones belong in which data
         structures should be separate duties, performed by separate
         servers.
  \item Incentives should be left to applications: servers can
         attest to any blocks they wish, for whatever reason.
  \item When reading a data structure, it should be possible to
         know the inherent trust assumptions: 
        Under what conditions will this data remain available?
        When might someone read conflicting data?
  \item Data should always be able to become \textit{more}
         trustworthy, and \textit{more} available, but never less so.
\end{itemize}

There are two types of servers in Charlotte: \textit{Wilbur}
 availability servers are key-value stores which store blocks, and
 \textit{Fern} integrity servers attest to which blocks belong in
 which data structures.
Together, these servers provide the {\blockweb}: a decentralized data structure with
 many smaller data structures within it.
Applications define the structures they wish to work with, and most
 importantly, specify who must attest to blocks in those structures.
These specifications effectively define the application's failure
 assumptions: if the application only specifies one specific server on
 which to store a block, it assumes that server won't crash.
Minting a block within an application involves creating a block with
 whatever data the application uses, and then sending it to servers
 that work with the application, so they can attest  that it is
 stored, and that they will not attest to conflicting blocks.
This process is shown in~\autoref{fig:life}.
Anyone reading the set of sufficiently available blocks and
 selecting those with the appropriate attestations will see all
 the blocks created for a given application (unless that application's
 failure assumptions were wrong).

  \subsection{Structure of a Block}
  \label{sec:structure}
  \begin{figure}
  \centering
  \resizebox{0.8\columnwidth}{!}{
\begin{tikzpicture}
\draw (4.5,6.5) node[draw, anchor=south]{\textbf{Key} (root hash)} -- (4.5,6.06) node[draw, anchor=north,text width=1.8cm, text height=.4cm]{};
\draw (0,0) -- (8.5,0) -- (8.5,6.2) -- (0,6.2) -- (0,0);
\node[align=left,anchor=west] at (0,6) {\textbf{Block}};

\draw (4,5.5) node[draw, anchor=south]{hash} -- (3.5,5.06) node[draw, anchor=north,text width=1.8cm, text height=.4cm]{};
\draw (5,5.5) node[draw, anchor=south]{hash} -- (7.6,3) node[draw, anchor=north,text width=1.45cm]{\textbf{Payload}\\application specific};

\draw (3,4.5) node[draw,anchor=south]{hash} -- (2,4);
\draw (4,4.5) node[draw,anchor=south]{hash} -- (5.4,4) node[draw, anchor=north,text width=1.8cm, text height=.4cm]{};

\draw (4.9,3.43) node[draw,anchor=south]{hash} -- (4.9,3) node[draw, anchor=north,text width=1.45cm]{\textbf{Payload}\\application specific};
\draw (5.9,3.43) node[draw,anchor=south]{hash} -- (5.9,1.5) node[draw, anchor=north, text width=1.8cm]{\textbf{Reference}\\ to another block};

\draw (.1,.1) -- (3.9,.1) -- (3.9,4) -- (.1,4) -- (.1,.1);
\node[align=left,anchor=west] at (0,3.8) {\textbf{Label}};

\draw (.2,2.2) -- (3.8,2.2) -- (3.8,3.5) -- (.2,3.5) -- (.2,2.2);
\node[align=left,anchor=west] at (.2,3.3) {\textbf{Availability Policy}};
\node[align=left,anchor=west] at (.12,2.7) {who will store this block \\ and how they'll prove it};

\draw (.2,.2) -- (3.8,.2) -- (3.8,2) -- (.2,2) -- (.2,.2);
\node[align=left,anchor=west] at (.2,1.8) {\textbf{Integrity Policy}};
\node[align=left,anchor=west] at (.15,.9) {who will guarantee no \\ conflicting blocks, and \\ how they guarantee it};
\end{tikzpicture}
}
\ICS{This does not look very good.
     Maybe it needs colors or symbols or something.}
  \caption{Structure of an Example Block}
  \label{fig:block}
\end{figure}
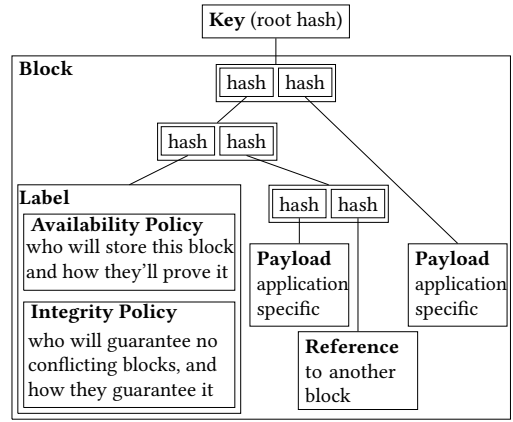

Each block is a Merkle Tree~\cite{merkle-trees}.
\ACM{Is it necessarily a tree? Presumably it could be a DAG.
}\ICSreply{At the moment, actually, it's defined to be a tree.
           That's pretty common for blockchains, since Ethereum.
           We could, in principle, alter the spec to allow DAGs.
           Is it worth it at this point?}
There are three leaf types:
\ACM{You don't say anything about the format of non-leaf nodes?
}\ICSreply{Being a ``Merkle Tree'' pretty much defines that.
           Our non-leaf nodes are encoded in Thrift as a map from hash
            to child nodes, and hashed as the hash of the
            concatenation of all the hashes which are keys to said
            map, ordered lexicographically.
           Is that worth saying?}
\begin{itemize}
\item \textit{Payload} leaves are application-specific data.
\item \textit{Reference} leaves identify another block, and provide
       attestations as to where the block is stored, and what data
       structures include it~\p{\autoref{sec:attestations}}.
\item \textit{Label} leaves describe where this block should be
       stored, what data structures it must be a part of, and who must
       approve its membership in those data
       structures~\p{\autoref{sec:labels}}.
\end{itemize}
Our implementation specifies the data layout of references and
 labels~\p{\autoref{sec:implementation}}.
The root hash of the tree is used as the \textit{key}, or unique
 identifier, of the block. 
Merkle proofs can demonstrate the presence of a label or
 reference in a block~\cite{merkle-trees}.
\autoref{fig:block} shows an example of a block.

Charlotte provides a data format for blocks and for references to
 blocks.
This structure is agnostic to implementation-specific details such as
 hash functions and digital signature formats, so
servers can specify the formats they use.

    \subsubsection{Attestations}
    \label{sec:attestations}
    \begin{figure}
  \centering
  \resizebox{0.8\columnwidth}{!}{
\begin{tikzpicture}
\draw (4.5,6.5) node[draw, anchor=south]{\textbf{Key} (root hash)} -- (4.5,6.06) node[draw, anchor=north,text width=1.8cm, text height=.4cm]{};
\draw (0,0) -- (8.5,0) -- (8.5,7.2) -- (0,7.2) -- (0,0);
\node[align=left,anchor=west] at (0,7) {\textbf{Reference}};

\draw (4,5.5) node[draw, anchor=south]{hash} -- (3.5,5.06) node[draw, anchor=north,text width=1.8cm, text height=.4cm]{};
\node[draw, anchor=south] at (5,5.5) {hash};

\draw (3,4.5) node[draw,anchor=south]{hash} -- (2,4);
\node[draw,anchor=south] at (4,4.5) {hash};

\draw (.1,.1) -- (3.9,.1) -- (3.9,4) -- (.1,4) -- (.1,.1);
\node[align=left,anchor=west] at (0,3.8) {\textbf{Label}};

\draw (.2,2.2) -- (3.8,2.2) -- (3.8,3.5) -- (.2,3.5) -- (.2,2.2);
\node[align=left,anchor=west] at (.2,3.3) {\textbf{Availability Policy}};
\node[align=left,anchor=west] at (.12,2.7) {who will store this block \\ and how they'll prove it};

\draw (.2,.2) -- (3.8,.2) -- (3.8,2) -- (.2,2) -- (.2,.2);
\node[align=left,anchor=west] at (.2,1.8) {\textbf{Integrity Policy}};
\node[align=left,anchor=west] at (.15,.9) {who will guarantee no \\ conflicting blocks, and \\ how they guarantee it};

\node[draw, align=left, anchor=east, text width=4cm] at (8.4,3.5) {\textbf{Attestation} (availability) \\ signed by Wilbur server};
\node[draw, align=left, anchor=east, text width=4cm] at (8.4,2.2) {\textbf{Attestation} (availability) \\ signed by Wilbur server};
\node[draw, align=left, anchor=east, text width=4cm] at (8.4,.9) {\textbf{Attestation} (integrity) \\ signed by Fern servers};
\end{tikzpicture}
\ICS{This does not look very good.
     Maybe it needs colors or symbols or something.}
}
  \caption{Structure of an Example Block Reference}
  \label{fig:reference}
\end{figure}
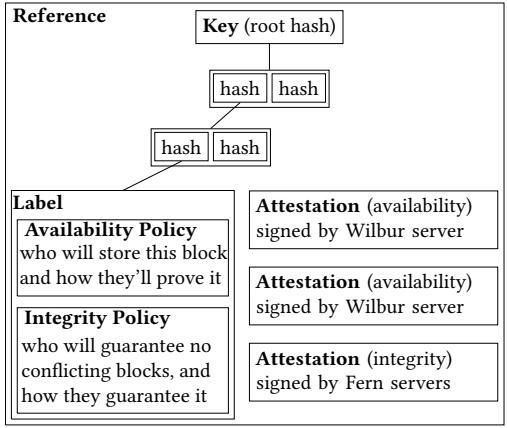

Each attestation is a promise.
Wilbur servers create \textit{availability attestations} which promise
 to store a block.
Fern servers create \textit{integrity attestations} which promise
 \textit{not to} attest to conflicting blocks.
Attestations can serve as part of evidence of wrongdoing, supporting
 applications' incentive systems.

Since attestations cannot be created until after the block is formed
 (for instance, a Wilbur server cannot store a block until after the
 block exists),
 attestations for a block cannot be stored in the block itself.
Therefore, each reference to a block must include any relevant
 attestations.
If a block references its predecessor in a chain, for example, it
 includes attestations demonstrating that the predecessor is
 sufficiently well stored, and uniquely occupies the previous chain
 slot.

Any reference to a block should demonstrate that
 \textit{at least} the block's own label is satisfied.
Therefore, a reference to a block should include a copy of the label,
 and a Merkle proof that the label corresponds to the block's root
 hash.
Additional attestations may bolster the block's demonstrated
 availability and integrity.
\autoref{fig:reference} shows an example of a block reference.
\ICS{Tom wonders if the preceeding paragraph really ought to use
      ``should,'' or ``must.''
     Technically, the answer is ``should.''
     Applications can be made to work just fine without ever actually
      checking attestations or labels.
     However, they'd lose any of Charlotte's guarantees.
     So, they ``should'' include these things.}

\paragraph{Availability Attestations}
 are issued by Wilbur servers.
One type, which we have implemented in our prototype, is a signed
 promise to store a block forever.
Charlotte is extensible with respect to proof types, so other types, such
 as proofs of retrievability~\cite{Bowers2009}, may be added by any
 servers that support them.

\paragraph{Integrity Attestations}
 are issued by Fern servers.
For example, if a Fern server in our supply chain application attests
 to a block stating that {\principalA} sold a specific unit of
 {\commodityA} to {\principalB}, the Fern server will not attest to
 any other blocks in which {\principalA} sells that unit of
 {\commodityA}\ICS{... unless she gets that unit back again. 
                   Is there a succinct and non-confusing way to work
                    that in?}.
One type of attestation, found in our prototype implementation, is a
 \textit{proof of consensus} (\autoref{sec:consensus}).
Consensus participants attest that they will not consent to
 conflicting blocks.
All messages in the consensus are signed, and the proof
 is a collection of the messages sent while achieving consensus.
Other conceivable integrity attestations include proofs of work.
Unlike a proof of consensus, where the underlying assumption is that a
 certain portion of the participants are honest, the underlying
 assumption of a proof of work attestation is that no more hashpower
 has been used on a different block.
When looking at all the blocks in a proof of work chain, the most
 trustworthy attestations are those demonstrating the most
 hashpower has been applied to a block (or its descendants).
This is a natural implementation of the \textit{longest chain}
 rule~\cite{bitcoin}.

\paragraph{The Charlotte framework} is \textit{extensible}:
 attestations are a tagged union type, so
 anyone can add application-specific attestation types
 as long as their servers know what to do with them.
In general, not all servers support all proof types.
Applications must be engineered only to ask for proofs from servers
 that can provide them.
Our prototype implementation supports queries and
 error messages for discovery of available proof types.
\ICS{
I have removed mention of the implementation specification.
This is for 2 reasons:
First, the specification, which is a block reference, is not
 inherently unique: it just kicks the problem of ambiguous parsing to
 the block referenced.
Second, any application minting a data type, which is worried about
 that type being misparsed, may include anything in the data type they
 wish, in order to indicate to servers what they ought to do.
This includes any manner of implementation specification.
The old implementation specification text is as follows:
To ensure different machines don't parse the same attestation two
 different ways, we allow each instance to include an
 \textit{Implementation Specification}, which can reference a unique
 block.
Applications can choose how exactly to canonicalize attestation types in
 such blocks.
}

\paragraph{Semantically,}
 data of type $\tau$ are availability attestations if $\tau$ defines a
 unique function $\alpha_\tau$ from a $\tau$ to a set of sets of
 Wilbur servers:
\[
\alpha_\tau : \tau \rightarrow  2^{\p{2^W}}
\]
(where $W$ is the set of all Wilbur servers)
The attestation $A$ must hold so long as one of the sets in
 $\alpha_\tau\p A$ is composed of Wilbur servers who are all behaving
 correctly.

Suppose a client has an attestation $A$, and queries one Wilbur server
 from each set in $\alpha_\tau\p A$.
If all the Wilbur servers respond ``block not found,'' then the client
 has proof of an attestation violation: no set in $\alpha_\tau\p A$ is
 composed entirely of correct servers.
Denial-of-service attacks in which the server simply does not respond
 are also possible, although there is work in creating proofs of those
 as well~\cite{Herlihy2016}.

Data of type $\tau$ are integrity attestations if $\tau$ defines two
 unique functions $c_\tau$ and $\alpha_\tau$ of types:
\begin{equation*}
c_\tau : \p{\tau \times \tau^\prime} \rightarrow \textrm{Boolean} \rule{0.6in}{0in}
\alpha_\tau : \tau \rightarrow 2^{\p{2^F}}
\end{equation*}
(where $\tau^\prime$ is a type of attestation,
 and $F$ is the set of all Fern servers)
The function $c_\tau$ defines which attestations \textit{conflict} with
 an attestation, and
 $\alpha_\tau$ defines the set of universes (in terms of which Fern
 Servers are trustworthy) in which an attestation is safe.

Suppose attestation $A$ is of type $\tau$, and attestation $A^\prime$,
 of type $\tau^\prime$, conflicts with $A$.
By definition, this means that $c_\tau\p{A,A^\prime}=True$.
The pair $\p{A,A^\prime}$ constitutes a proof that every set of
 Fern servers in $\alpha_\tau\p{A}$ contains a byzantine server.

We call attestation $A$ \textit{stronger} than attestation
 $A^\prime$ when $A$ conflicts with a superset of the attestations
 $A^\prime$ conflicts with, and $A$ holds whenever $A^\prime$ holds:
\begin{equation*}
\forall x .   c_\tau\p{A,x}             \Rightarrow c_\tau\p{A^\prime,x} 
 \rule{0.6in}{0in}
              \alpha_\tau\p{A^\prime}   \subseteq \alpha_\tau\p A
\end{equation*}

\ICS{I have removed a bit about how these semantics correspond to our
 implementation.
It went like this:
In our implementation, each data type $\tau$ must also be expressible
 in the Thrift rpc language\ICS{citation needed}, and we identify
 servers by public key.
Our prototype servers have literal instantiations of the functions
 $\alpha_\tau$ and $c_\tau$ in Haskell.
}

Correct servers do not issue attestations that conflict with
 attestations that others might make.
For instance, our proof of consensus attestations are constructed so
 that byzantine failures too weak to influence the consensus that
 created that attestation cannot create conflicting attestations.
\ICS{This section is now very long.}

    \subsubsection{Labels}
    \label{sec:labels}
    Each block contains exactly one \textit{label} leaf.
This leaf describes which data structures the block is meant to be a
 part of, and what is necessary for its approval.
Specifically, a label describes the least strong attestations a block
 requires.
In some sense, if attestations are never created satisfying the
 block's label, the block has \textit{failed}, and should not be
 referenced.
For instance, in our supply chain example~\p{\autoref{sec:supply}},
 each block representing a trade would have to be approved by the
 consensus mechanisms tracking the goods involved.
Each consensus mechanism is a consortium of distinct principals, which
 agree together to attest to blocks representing supply chain
 transactions.
Additionally, those consortia might demand that any block they make a
 permanent part of the record must be stored in a certain way.
Thus each block's label requires that it be stored in a particular way,
 and that it be approved by a particular consensus.
    \subsubsection{Availability Policies}
    \label{sec:policies-availability}
    Within a label, availability policies describe possible sets of Wilbur
 servers which must store a block, as well as what kind of attestation
 each must provide.
One type of policy, found in our implementation,
 is a set of acceptable sets of Wilbur servers, each identified by
 public key, as well as a specified format for each server's
 attestation.
    \subsubsection{Integrity Policies}
    \label{sec:policies-integrity}
    Just as availability policies specify the minimal availability
 attestations a block requires, integrity policies specify the minimal
 integrity attestations.
An integrity attestation is a statement about what \textit{won't} be
 attested to in the future, so an integrity policy is likewise a
 statement about what anyone who attests to this block shouldn't
 attest to later.
In short, it defines the set of blocks which conflict with this one.

By specifying conflicting blocks, integrity policies specify the data
 structures to which a block will belong.
If a data structure is non-exclusive, any block can claim membership:
 no integrity policy is required.
Many data structures, however, require exclusivity: admitting some
 blocks precludes others.
For instance, a chain consists of a series of slots, each of which may
 be claimed by only one block.
Any two blocks claiming the same slot on the same chain conflict.

Integrity policies must also specify the \textit{type} of attestation
 they require~\p{\autoref{sec:attestations}}.
This may include a proof of work, or, as in our implementation, a
 proof of consensus.
\ICS{not sure what to say here.
     The appropriate place for this discussion is in the attestations
      section.}

  \subsection{Consensus}
  \label{sec:consensus}
  For a consensus mechanism to be useful in Charlotte, it must be able
 to produce a \textit{proof of consensus},
 which is any data type,
 verifiable to an external observer, demonstrating that a specific
 consensus decided on a specific block.
In order for a proof of consensus to make any sense, it must specify
 its participants.
A reader must be able to determine under what failure conditions (in
 terms of which Fern servers might fail), a contradictory proof of
 consensus might be forged.
A proof of consensus can thus form an integrity attestation.
A corresponding integrity policy would specify the participants who
 must achieve consensus.

In general, we want Charlotte blocks to atomically join (or fail to
 join) all their data structures together.
The integrity policy should require a single, atomic consensus
 mechanism that satisfies the constraints of all the data structures
 to which the block belongs.
For example, in our supply chain application~\p{\autoref{sec:supply}},
 suppose a block details an exchange of {\commodityA} for
 {\commodityB}.
It would be awkward if the data structure tracking {\commodityA}
 accepted the block, but not the data structure for {\commodityB}.
When two integrity policies are combined into a single policy, such
 that all attestations satisfying the unified policy satisfy both
 original ones, we call that a \textit{meet}~\p{\autoref{sec:joins}}.

For consensus-based integrity policies to have a meet, the consensus
 mechanism must have a way to bring together the participants of both
 sub-policies.
This meet consensus mechanism should only be able to issue
 contradictory proofs of consensus when the consensus mechanisms
 of \emph{both} sub-policies are likewise compromised.  
\ICS{this is really awkward, as we formally define meets later...}

One such mechanism, provided in our prototype, is a
 byzantine Paxos-based consensus description, where each data
 structure dictates quorums required for approval, and the block can
 only be approved by a set of Fern servers including a quorum from
 each data structure~\cite{paxos-made-simple}.
Other consensus mechanisms, such as
 Stellar~\cite{mazieresstellar}, have a
 concept of
 \textit{meet} as well.
We are not aware, however, of another consensus implementation that
 creates a proof of consensus.

  \subsection{Structure of a Reference}
  \label{sec:reference}
  Each \textit{block reference} features the block's root hash, which
 uniquely identifies the block.
It also includes a set of attestations, allowing anyone retrieving
 the block to know where the block may be found, how available it is,
 and which data structures it is part of.
Finally, it features the block's label~\p{\autoref{sec:labels}}, and a
 Merkle proof that the label matches the root.
\autoref{fig:reference} is a diagram of a block reference.

  \ICS{Matthew suggests that the ``Life of a Block'' subsection is
        wholly redundant with preceeding material, and so it has been
        commented out.}
  \subsection{Chains}
  \label{sec:chains}
  \begin{figure}
  \centering
  \begin{tabbing}
  \rule{1.75in}{0in}\=\kill
  \begin{minipage}{2in} \newcommand{\hofffffset}[0]{0.34} 
\newcommand{\vofffffset}[0]{0.2}
\definecolor{babyblue}{RGB}{157,227,255}
\newcommand{\blueblockcolor}[0]{babyblue}
\newcommand{\refnode}[1]{\draw[line width=\reflinewidth, ->] (#1) node[circle, fill=black, text width=\refdiameter]{}}
\newcommand{\labeledblock}[5]{{
\node[draw, line width=0.6mm, anchor = north, text width=8mm, text height = 6mm, fill=#3] (#4) at ({{#1}},{{#2}}) {};
\node[below] at (#4.north) {#5};
\node[] (#4-l) at ($ (#4) -  (\hofffffset , \vofffffset) $){};
\node[] (#4-r) at ($ (#4) +  (\hofffffset,-\vofffffset) $){};
\node[] (#4-ul) at ($ (#4) -  (\hofffffset , \vofffffset - 0.25) $){};
\node[] (#4-ur) at ($ (#4) +  (\hofffffset , 0.25 - \vofffffset) $){};
\node[] (#4-cr) at ($ (#4) + (5 * \hofffffset/12, -\vofffffset) $){};
\node[] (#4-cl) at ($ (#4) - (5 * \hofffffset/12,\vofffffset) $){};
\node[] (#4-c) at ($ (#4) - (0,\vofffffset) $) {};
}}
\newcommand\refdiameter{-1.5mm}
\newcommand\reflinewidth{0.3mm}
\newcommand\bluechainx{4}

\resizebox{0.85\columnwidth}{!}{
\begin{tikzpicture}

\tikzmath{
    \vertsep = 2.0;
    \offl = .2; \offll = .4; \offv = 0.75;
    \refoffv = -1.3;
    \vertiv = 4*\vertsep - 0.6;
    \vertiii=3*\vertsep-0.1;
    \vertiiidot = \vertiii+\refoffv-0.4;
}

\labeledblock{0}{0}{red}{redR}{root}

\labeledblock{0}{1}{red}{red1}{1}
\refnode{red1-c} -- (redR);
\refnode{red1-l} -- ($ (redR.north) + (-0.2,0) $);

\labeledblock{0}{2}{red}{red2}{2}
\refnode{red2-c} -- (red1);
\refnode{red2-l} to [out=230, in=110] ($ (redR.west) + (0,0.4) $);

\labeledblock{\bluechainx}{1}{\blueblockcolor}{blueR}{root}

\labeledblock{\bluechainx}{2}{\blueblockcolor}{blue1}{1}
\refnode{blue1-c} -- (blueR);
\refnode{blue1-r} -- ($ (blueR.north) + (0.2,0) $);

\labeledblock{\bluechainx / 2 }{2.6}{red}{shared-color-only}{};
\node[ anchor = east, text width=3mm, text height = 6mm, fill=\blueblockcolor]  at (shared-color-only.east) {};
\labeledblock{\bluechainx / 2 }{2.6}{none}{shared}{3 2};
\refnode{shared-ul} -- (red2);
\refnode{shared-ur} -- (blue1);
\refnode{shared-cl} -- (redR.east);
\refnode{shared-cr} -- (blueR.west);

\labeledblock{0}{3.2}{red}{red4}{4}
\refnode{red4-r} -- (shared);
\refnode{red4-l} to [out=240, in=110] (redR.west);

\labeledblock{\bluechainx}{3.2}{\blueblockcolor}{blue3}{3}
\refnode{blue3-l} -- (shared);
\refnode{blue3-r} to [out=300, in=70] (blueR.east);

\end{tikzpicture}
} \end{minipage}
  \> \begin{minipage}{1.5in}
  \caption{Two blockchains sharing a block.
           The label of the shared block is the meet of the chains' labels.}
  \label{fig:chains}
	\end{minipage}
  \end{tabbing}
  \vspace{-8mm}
\end{figure}
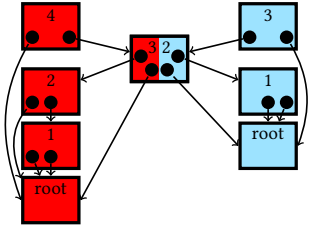

Many Charlotte applications, including in our prototype
 implementation, will wish to implement a blockchain.
A block\textit{chain}, broadly defined, is any path through the
 \blockweb.
However, it is convenient to specify a few more qualities
 usually implied by blockchains, and a natural way to
 represent them in Charlotte.
This is what we mean by \textit{blockchain}:

A blockchain can be identified by a unique root block.
This is simply the block in the chain that has no
 predecessors.
A chain must also have a label representing how blocks in the chain
 must be stored, and what attestations are required for the uniqueness
 of each block.
The natural label to use is the label of the root itself.
Thus the root satisfies the requirements of the blocks in the chain,
 and whoever issues attestations for the root acknowledges that the
 chain exists when attesting to the root.

Blocks can be in multiple chains.
Each block in a chain references both the previous block and the root
 of the chain.
It must also have a label that is at least as strong as that of each
 of its roots.
Integrity labels and attestations can be application-specific, so it
 is up to Fern servers to verify that they understand the labels
 involved, and have checked their relative strength.
Furthermore, as chains are totally ordered, a block should specify
 what position it takes in each chain, in terms of ``distance from the
 root.''
We call this distance a \textit{slot}, and integrity attestations for
 chains vow not to approve any two block with the same slot on the
 same chain.
\autoref{fig:chains} features two blockchains sharing a block.

In general, chains do not fill later slots before earlier ones.
A natural representation is to use natural numbers as slots
 (number of blocks from root).
By requiring each block to reference its immediate predecessor (and
 valid references include attestations that the predecessor is
 unique), we can guarantee that earlier slots are filled before later
 ones.
\ICS{The following has been removed from this section, as it is more
      about chain properties than design:
We say this guarantee is provided \textit{up to} the integrity of the
 chain's (root's) label.
Specifically, if the attestations on a block in the chain are
 \textit{violated}, meaning an attestation has been made for a
 conflicting block, then the chain properties fail.
The conditions under which this can happen are bounded by the chain's
 label:
If, for example, the chain requires a specific consensus to attest to
 each block, then the chain's properties will be violated only if the
 consensus attests to conflicting blocks, which can happen only if
 enough consensus participants are byzantine.
}

  \subsection{Entanglement}
  \label{sec:entanglement}
  Entanglement occurs when different applications interact.
Because each block contains a hash that uniquely identifies its
 ancestry, attempts to change old blocks become harder to
 cover up when they have more diverse descendants.
If, for example, many applications' blocks reference old timestamps
 (to show they were created after a given time), and in turn are
 later timestamped, then after a while, all these applications' new
 blocks would be descendants of all these applications' old blocks.
Their chains become \textit{entangled}.
It would be impossible for an adversary to change the old records of
 one application without changing records of all the applications.
  \subsection{Incentives}
  \label{sec:incentives}
  Charlotte does not specify \textit{why} Fern or Wilbur servers choose
 to attest to certain blocks.
This is, in general, specific to an application.
For instance, if a company runs a blockchain for their own use, they
 may run their own Fern and Wilbur servers, which operate over all
 blocks signed by company software.
It is also possible to run for-pay Fern or Wilbur servers.
For instance, a server might choose only to attest to blocks that
 include in their payload a transfer of funds to the server's owner
 (see \autoref{sec:banking} for how one might implement funding
 transfers).
We hope to see many incentive schemes implemented in Charlotte.

\section{Consistency Properties}
\label{sec:consistency}
Most existing blockchains rely on the serializable
 consistency~\cite{Papa79} of their transactions for critical
 properties~\cite{bitcoin,ethereum}.
The programming model offered by serializable transactions has proven
 attractive to Blockchain users.
With no requirement for total serialization, Chartlotte does not
 inherently have this property.
However, Charlotte can, in general, guarantee
 \textit{causal consistency}~\cite{Ahamad1995,COPS,Tardis}.
Furthermore, subgraphs of Charlotte, such as chains described
 in~\autoref{sec:chain}, can guarantee serializable consistency.
This is true even for data spanning multiple such chains.

Moreover, we can demonstrate that, unlike most permissionless
 blockchains, data structures in Charlotte maintain availability and
 integrity properties specified in their labels, reliant only on the
 Wilbur and Fern servers they trust\footnote{
   It is possible to specify permissionless data structures in
    Charlotte.
   In this case, the notion of who the structure ``trusts'' is less
    explicit than in most permissioned structures.
  \ICS{Not sure what, if anything, to say about this.}
 }.
\textbf{Untrusted adversaries can have no effect on them.}

  \subsection{Joins and Meets}
  \label{sec:joins}
  To describe Charlotte's consistency properties, it is useful to
 consider the possible universes in which the system may be operating.
For our purposes, each \textit{universe} is characterized by the set
 of servers which have not violated their attestations.
For instance, if we are living in a universe where all the servers
 have violated their attestations, then none of the guarantees implied
 by block labels hold, since no attestations can be trusted anyway.

Intuitively, availability attestations describe the set of possible
 universes under which a block is still available.
Likewise, integrity attestations describe the set of possible
 universes in which no conflicting block can be endorsed by equivalent
 attestations.
For an attestation $A$, we refer to this set of universes as
 $\alpha\p A$.

For instance, if a Wilbur server $W_1$ attests to a block $B$, then in
 all universes where $W_1$ holds true to its attestation, $B$ remains
 available.
Likewise, if a Fern server $F_1$ attests to a block $B$, then in all
 universes where $F_1$ is honest, no block conflicting with $B$ will
 have an attestation from $F_1$.

We consider an attestation to be \textit{stronger} than another if it
 describes a superset of possible universes, and equivalent if it
 describes the same set.
Conversely, a stronger attestation fails in a subset of the universes
 in which a weaker attestation fails.
When a block has multiple attestations, we can consider their
 \textit{join} \p\join: a \textit{compound attestation} representing
 the universes in which at least one of the attestations holds.
\[
\alpha\p{\bigsqcup A} := \bigcup_{a\in A} \alpha\p a
\]
(sometimes written $a\join a^\prime := \bigsqcup \cb{a,a^\prime}$ for
 two attestations)
\ACM{Giving the syntax of meet and the semantics of meet in one
      formula -- too much happening at once here.
     It's not clear to me that the join/meet notation is helping yet.
     Maybe just use the set notation and introduce join and meet if
      and when they are needed to express things clearly?
     Presumably this is when talking about a policy language?
}\ICSreply{It's actually not something we use a lot at this time.
           I think, however, that we might actually want to use it
            more, since both Andru and Robbert agree that we could use
            more technical detail.
           We should be able to say that the \meet of a data structure
            is the \meet of all the attestations on all the blocks
            therein, which is bound to be at least as strong as the
            \meet of all the labels of the blocks therein.
           Likewise, we should be able to say that the label of a
            block on multiple chains should be at least as strong as
            the \join of the labels of each chain.
           This section gives us a place to refer for those technical
            definitions.
           I'm not sure where else to do it elegantly.
}
Thus we consider one set of availability attestations $A$ stronger than
 another set $A^\prime$ when the join of $A$ is stronger than the join
 of $A^\prime$.
Likewise, we consider one set of integrity attestations $I$ stronger
 than another set $I^\prime$ when the join of $I$ is stronger than the
 join of $I^\prime$.

We can also consider the compound attestation representing the set of
 universes where all of the attestations in a set hold.
We call this the \textit{meet} \p{\meet} of the set:
\[
\alpha\p{\bigsqcap A} := \bigcap_{a\in A} \alpha\p a
\]
(sometimes written $a\meet a^\prime := \bigsqcap \cb{a,a^\prime}$ for
 two attestations.)

For blocks, we want to consider the set of universes in which they are
 both available \textit{and} no conflicting blocks have equivalent or
 stronger attestations.
For a block $B$ with availability attestations $A$, and integrity
 attestations $I$:
\[
  \alpha\p B := \p{\bigsqcup A} \meet \p{\bigsqcup I}
\]

For instance, suppose \principalA{} and \principalB{} are minting a
 block together.
They get availability attestations from Wilbur servers $W_1, W_2$, and
 $W_3$, and integrity attestations from Fern servers $F_1, F_2$, and
 $F_3$.
In all universes where $W_1, W_2$, \textit{or} $W_3$ are live and
 trustworthy, the block will remain available.
Furthermore, so long as $F_1, F_2$, or $F_3$ are trustworthy, no
 conflicting block can get equivalent or stronger attestations.
When we consider $\alpha$ for proofs of work, each universe is still
 characterized by the set of trustworthy attestations: in this case,
 the set of proofs such that no equal or longer conflicting proof of
 work exists.
Thus longer proofs of work are equivalent to more attestations.
\ICS{This is not well explained.}

Blocks with a superset of attestations are more available, or more
 trustworthy, than other blocks.
For instance, suppose block $B_1$ is endorsed by Fern server $F_1$,
 but  block $B_2$ is endorsed by Fern servers $F_1$ and $F_2$.
If $F_1$ is dishonest, then blocks conflicting with $B_1$ might also have
 $F_1$ attestations: no client would have reason to prefer $B_1$ over
 those conflicting blocks.
Therefore, we say $B_2$ has \textit{stronger} integrity, because only
 when $F_1$ and $F_2$ are dishonest can $B_2$ be contradicted by a
 block with equivalent attestations.
Note that many attestations are \textit{incomparable}: if
 $B_3$ has attestations from Fern servers $F_2$ and $F_3$, then should
 $B_2$ and $B_3$ conflict, Charlotte sets no precedent as to which is
 ``preferable.''

\ACM{This paragraph is where we should first talk about joins and meets.
     (we used to start this subsection with 
     ``To describe Charlotte's consistency properties, it is useful to
        be able to talk about \textit{join}~\p\join{} and
        \textit{meet}~\p\meet.'')
}\ICSreply{I did move the definition down, but I also reworked it to
            be more precise, and define join and meet over
            attestations.
           As a result, the introduction of the terms and symbols is
            not quite this far down.
           I hope that's OK.
}
When considering multiple blocks, we can talk about the set of
 universes in which both are available and uncontradicted.
This is precisely when at least one availability attestation of each
 features an honest and live Wilbur server, and at least one integrity
 attestation for each is from an honest Fern server.
Thus the \textit{meet} $\p\meet$ of a set of blocks $B$ is defined
 exactly like the meet of a set of attestations.

Likewise, we can talk about the set of universes where at least one
 block is available and uncontradicted.
Thus the \textit{join} $\p\join$ of a set of blocks $B$ is defined
 exactly like the join of a set of attestations.

Join and meet for references are defined exactly as for blocks.
We can even define $\alpha$, join, and meet for labels.
For a label $\ell$, let $R$ be the set of all possible references
 featuring attestations satisfying $\ell$:
\[
\alpha\p\ell := \alpha\p{\bigsqcap R}
\]
Intuitively, the set of universes in which a label is violated includes
 any universe where any reference satisfying that label is violated.
Join and meet are then defined for labels exactly as they are for
 blocks and references.

  \subsection{Causal Consistency}
  \label{sec:causal}
  Causal consistency, intuitively, is the requirement that in any view
 of a system, if an effect of an event is visible, then so are the
 effects of any logical causes of that
 event~\cite{Ahamad1995,COPS,Tardis}.
In Charlotte, we think of each block as an event, and each block it
 references as a logical cause.
Each reference carries attestations that the blocks referenced are
 available and approved members of their respective data structures.
The DAG structure of the {\blockweb} guarantees no cyclic causality.
At a very basic level, causal consistency is assured.

However, in Charlotte, the notion of event ``visibility'' is not so
 clear-cut.
Different blocks may have different levels of availability and
 integrity.
When, for instance, one set of Wilbur servers crash, some blocks may
 become unavailable, but not others.
Therefore, we define two notions:
\begin{itemize}
\item The \textit{recursive availability} of a block is the
       availability of that block, joined with the recursive
       availability of the blocks it references.
      This may be thought of as the set of circumstances (in terms of
       which Wilbur servers hold to their attestations) under which
       the block's entire ancestry is available.
\item The \textit{recursive integrity} of a block is the integrity of
       the block, joined with the recursive integrity of each block it
       references.
      This may be thought of as the set of circumstances (in terms of
       which integrity attestations are violated) under which no block
       is approved, which conflicts with any block in the ancestry.
\end{itemize}
Thus, \textbf{the causal consistency of any block
 is assured precisely when the block's recursive availability and
 recursive integrity hold.}

As time goes on, and ancestries of new blocks grow larger, it may seem
 that recursive integrity and recursive availability of blocks would
 naturally become weaker.
However, each time a new block is approved, Fern and Wilbur servers
 may add new attestations to those it references.
In fact, some applications may specify
 \textit{recursive attestations}, which are a short-hand for listing
 an attestation for each block in the given block's ancestry.
For instance, a proof of work attestation demonstrates that a certain
 amount of hash-reversing power is necessary to approve a conflicting
 block.
Because the block contains hash roots of those it references, that
 amount of hash-reversing power would also be necessary to approve
  blocks conflicting with those referenced.
Since additional hash-power is added each time a new descendant is
 minted, it is possible to construct ever-stronger attestations of old
 blocks as time goes on.

If an application considers its Fern servers to implicitly approve a
 block's ancestry each time they approve a block, then a block's
 recursive integrity is exactly its own integrity.
Furthermore, if an application stores all elements of a block's
 ancestry in the same way as it stores the block itself, then the
 block's recursive availability is the same as its availability.
If an application does both, then a block's causal consistency is
 assured whenever the block's labels hold.

Charlotte does not have an explicit requirement that all attestations
 are recursive in this way, because some applications may not benefit
 from such a thing~\p{\autoref{sec:timestamping}}.
Furthermore, some applications may consider some historical blocks
 more ``important'' than other historical blocks. 
For example, with the timestamping
 application~\p{\autoref{sec:timestamping}}, blocks referencing a
 timestamp have all manner of unrelated ancestors, since timestamp
 blocks can reference blocks from any application.
For such applications, the relevant consistency properties will have
 to do with recursive availability or integrity \textit{restricted to}
 the set of blocks they consider important.

Causal consistency alone is sufficient for some applications.
\ICS{It may be useful here to cite an application for which causal
      consistency is sufficient.
     I would cite timestamping, but we just finished explaining how
      timestamping actually gets blocks extremely unreliable recursive
      integrity and availability.
     Email would be a good example, but it's not in our applications
      section.}
However, other applications may want the full serializability provided
 by traditional blockchains.

  \subsection{Chain Properties}
  \label{sec:chain}
  When a chain is constructed according to~\autoref{sec:chains}, blocks
 have a serializable consistency property~\cite{Papa79}.
In all universes specified by the label of the root block of the
 chain, the blocks have a serial order defined by their slots.
We say that a property holds \textit{up to} a label when it holds in
 all universes described by that label.
Therefore, we say that blockchains are serializable up to the label of
 the chain.

For every block $B$, all other blocks sharing a chain with $B$ are
 serialized either before $B$ (they are ancestors of $B$), or after
 $B$ (they are descendants of $B$).
This property holds up to the meet of the labels of the chains $B$ and
 the other block share.
That is to say, if $B$ and $B^\prime$ are both on a chain $C$, and
 Fern servers have not issued two contradictory attestations for $C$,
 then $B$ and $B^\prime$ are ordered.
Since the attestations of all blocks in a chain are bounded by the
 chain's label, the blocks of a chain are serialized up to the label
 of the chain.

If a set of blocks share multiple chains, they are serialized so long
 as one of the chains they share is serialized.
A set of blocks is serialized up to the join of the labels of the
 chains they all share.

    \subsubsection{Chains as Objects}
    \label{sec:chains-as-objects}
    We can view each chain as a stateful object. 
Each block represents achange to that state.
At the lowest level, the object's state is simply the chain itself,
 but applications may parse it to something more useful.
For example, a chain representing an account balance could list
 balance changes in each block, and an application might read the
 account balance as the sum of all the account changes.

When all the blocks in the chain are available, the object's state can
 be said to be available.
Therefore the object's availability is bounded by the meet of the
 blocks in the chain, which is in turn bounded by the label of the
 chain.
The object is at least as available as the label of the chain
 indicates.

Likewise, when no attestation has been improperly issued for a block on
 the chain (such as if there's a conflicting block in the same slot),
 the object's state is consistent: all readers will see the same
 chain.
The object's integrity is therefore bounded by the label of the chain.

    \subsubsection{Blocks as Transactions}
    \label{sec:blocks-as-transactions}
    Each block can be on multiple chains.
In this object model, each transaction reads data from some chains,
 and writes data to some chains.
We represent each transaction as a block that is appended to all the
 chains it reads \textit{or} writes.
In this model, \textbf{our design for blockchains within
 Charlotte corresponds exactly to a serializable, atomic transaction
 system,} a popular programming abstraction~\cite{Papa79,gray93}.

Each block can be seen as ``locking'' all relevant chains, forcing all
 other operations to be scheduled either before or after the block.
The label of a chain bounds the object's availability, and integrity
 (an object loses integrity if it loses serializability).
In this light, Charlotte is a generalization of distributed,
 federated object stores~\cite{jfabric,oceanstore}\ICS{
   What else should I cite here?}.

\section{Applications}
\label{sec:applications}
Charlotte facilitates more flexible application designs than existing
 blockchain systems can support.
More Charlotte applications are described in
 \autoref{sec:additional}.
\ICS{Not sure what to say, if anything, here.}

  \subsection{Simulating Existing Chains}
  \label{sec:simulating}
  Charlotte is generic enough to faithfully emulate almost all existing
 blockchains and blockchain-like systems.
For instance, one could implement ``CharlotteBitcoin'' with blocks
 featuring the same payload as {\bitcoin} blocks, and proofs-of-work as
 integrity attestations.
Bitcoin's peer-to-peer block distribution network provides, in effect,
 very weak availability attestations.
CharlotteBitcoin could hold Wilbur servers accountable when they do
 not store blocks properly, since availability attestations can be
 used as proof of wrongdoing, and incentive systems could punish
 misbehaving servers.

Systems like \blockchainName{Iota}~\cite{iota},
 \blockchainName{Nano}~\cite{nano}, and
 \blockchainName{Spectre}~\cite{Sompolinsky2016} do not each strictly
 keep \textit{a chain} of blocks.
Nevertheless, their block payloads could
be formatted as Charlotte
 blocks, and reference multiple priors.
\ICS{We used to ahve a bit about interoperability here.
     Andru did not like it, so we removed it.
     Now I suppose this section begs the question, 
      ``why would you want to simulate other chains?''
     The only real inherent advantage I can think of is that there
      would at least be a format for referencing blocks in other
      chains, or being on two chains, so, interoperability.}


  \subsection{Object Model}
  \label{sec:object}
  As discussed in~\autoref{sec:chain}, we can think of chains in
 Charlotte as stateful objects, and each block as an atomic
 transaction.
The labels of the chains represent availability and integrity
 guarantees for the objects.
Blocks, in this light, represent serializable transactions.
This can be a helpful model for Charlotte applications.

\ICS{Not sure if the following paragraph is necessary}
As in \ethereum, it is possible to place programs and object state
 digests in each block.
While this is sometimes useful, it is not always necessary.
So long as the application can derive the state of an object from the
 payload of blocks on its chain, and each transaction is a block on
 all the chains it reads or writes from, the transactions are
 serializable.

    \subsubsection{Supply Chain}
    \label{sec:application-supply}
    Our Supply Chain example application~\p{\autoref{sec:supply}} features
 an object for each instance of an organization tracking a type of
 commodity.
Each block details a transfer of ownership, creation, or destruction
 of a commodity.
Likely, one would want to have each block signed by any commodity
 owners involved.

When \principalC{} wants to trade \coffee{} to \principalD{} in
 exchange for \diamonds{}, they execute an atomic transaction on the
 objects maintained by mutually agreeable commodities trackers.
Suppose these are the Canadian Coffee Catalogue (CCC), and the Danish
 Diamond Directory (DDD).
The transaction should thus be logged on both the CCC's object, and
 the DDD's object.
    \subsubsection{Conflicts \& Perceived Serializability}
    \label{sec:disreputable}
    Any party wishing to maintain a serializable view of the world need
 only commit every transaction it cares about to its own object, after
 examining the transaction's ancestry for conflicts with known
 history.
For example, suppose that \principalC{} and the CCC are malicious, and
 \principalC{} logs many sales of \coffee{}, citing as her source each
 time the same unit, which she purchased only once.
After \principalD{} logs the \coffee{} purchase with the DDD, the
 DDD's records link \principalC{}'s \coffee{} purchase with
 the sale of that \coffee{} to \principalD{}.
Assuming the DDD is appropriately checking the history of the goods in
 the transactions it commits, it will not accept any other
 transactions featuring a different sale of the same unit of \coffee{}
 anywhere in their history.
From the DDD's perspective, that \coffee{} was only sold once.
The serializability safety property is assured, from the DDD's
 perspective.
If the CCC attests to other transactions in which that \coffee{} was
 sold other times, the DDD simply will not accept any child blocks of
 those transactions.

Accepting transactions from untrustworthy sources (like the CCC) can
 thus be a danger to liveness.
To avoid locking itself out of potential future transactions (in
 essence having a different view of \coffee{} history from the other
 commodities traders) it is in the DDD's best interest to share any
 transactions it knows about, and get other principals to attest to
 them.
This should not be surprising: spreading the news of a sale helps to
 prevent double-selling.
For a similar reason, existing blockchain systems often gossip blocks
 between storage nodes~\cite{bitcoin}.
Charlotte provides the means to do so (servers can submit
 blocks to other servers, just like anyone else), but does not
 bake in any particular gossip requirement.
Applications are free to implement whatever best suits their needs.
Even when a malicious server successfully attests to two conflicting
 transactions, anyone discovering the two possesses evidence of
 duplicitousness, and an incentive system (such as law enforcement)
 may be able to limit the damage.

  \subsection{Banking}
  \label{sec:banking}
  Many existing blockchain applications implement currency transfers,
 the traditional domain of
 banks~\cite{bitcoin,ethereum,mazieresstellar,Schwartz2014}.
\ICS{I'm not sure of exactly how many citations we should put here.}
Charlotte's object model~\p{\autoref{sec:object}} provides a natural
 implementation for banking.
\ICS{Should this be a subsubsection of the object model subsection?}
Each bank account is an object, with a label dictating what Wilbur
 servers the account owner trusts to keep the account available, and
 what servers the account owner trusts to attest only to
 non-fraudulent transfers.
The label describes exactly who the account owner trusts to be a bank.
The advantage over traditional banking, however, is that account
 owners do not have to place all their trust in one organization.
For example, an account might set its label to be the
 meet~\p{\autoref{sec:joins}} of a label representing an account at
 one bank, and a label representing an account at another bank.
Then the account can be defrauded only if two banks are fraudulent
 (at the expense of losing the ability to make transactions if either
 bank crashes).
In fact, anyone can provide Fern and Wilbur servers, and account
 owners are free to set whatever label parameters they trust,
 making it easy to effectively create new banks.
\ACM{Not sure what the point of this sentence is here.
}\ICSreply{Any better now?}

\ICS{I have commented out a paragraph, because
     everything in this paragraph is done in more detail in the
      evaluation section.
I have commented out another paragraph because everything in the
      paragraph repeats stuff
      we said in the object model, just with a banking focus.}

  \subsection{Timestamping}
  \label{sec:timestamping}
  While the authenticated DAG structure of the {\blockweb} provides
 proof of whether blocks were created before or after other blocks, a
 \textit{timestamping} service provides evidence of whether a block was
 created before or after a given clock time.
Our timestamping application is an example of an application outside
 the Object Model.
It can take greater advantage of Charlotte's decentralized features.
In fact, it does not explicitly form blockchains at all.

A timestamping server simply issues a signed block stating the present
 time, and referencing other blocks as ``before that time.''
The timestamp's availability is set by availability attestations, but
 it needs no integrity attestation.
Any block the timestamp references is, if the timestamper is to be
 believed, provably created before that time.
Likewise, any block referencing the timestamp was created after that
 time.
Blocks can reference or be referenced by blocks issued by multiple
 timestamping servers, to increase the trustworthiness of their timing
 claims.

\subsubsection{Transitivity}
Timestamping is transitive.
Any ancestors of a timestamp are before that time, and any descendants
 are after it.
It is possible to construct a
 \textit{recursive attestation}~\p{\autoref{sec:attestations}} for any
 ancestor of a timestamp block.
The timestamp server attests that none of them will have as ancestors
 a timestamp block with a later time (from the same server).

Each timestamp block may have limited space to reference all the
 blocks which want to be timestamped.
Blocks willing to have a slightly later timestamp may agree to be
 referenced by another block, in effect a timestamp reseller,
 which is then referenced by the timestamp.
If applications regularly arrange for their blocks to be descendants
 and ancestors of timestamps (preferably multiple timestamps), then
 Charlotte as a whole will gather a high degree of
 \textit{entanglement}~\p{\autoref{sec:entanglement}}.
\ICS{I have cut the following, because it's already said in composability:
       Servers may not want to attest to the
        \textit{recursive integrity} or
        \textit{recursive availability}~\p{\autoref{sec:causal}} of
        blocks' timestamp-based ancestors, as these will include all kinds of
        unrelated applications.
       This is one of the reasons Charlotte does not require that all
        attestations be recursive.}

\subsubsection{Timestamping Implementation}
We implemented our timestamping application on top of 4 Wilbur
 servers.
Each server issues a timestamp block whenever it has stored 10 new
 blocks.
It then sends the new timestamp block to all the other servers.
A single client sent thousands of unique blocks to each server, and we
 timed how long it took for each to become the ancestor of timestamp
 blocks from at least 3 timestamping servers.

We compare to a more traditional blockchain, which completely
 serializes all blocks.
This also provides a kind of timestamping, as each round of consensus
 includes a timestamp endorsed by the participating Fern servers.
Indeed, existing services use existing blockchains as reliable
 timestamps~\cite{Gipp2015,origin-stamp,open-timestamps}.
In our experiment, each block was approved by a 3-out-of-4 byzantine
 tolerant consensus, and we timed how long it took to append each
 block to the chain:

\ 

\noindent
\begin{tabular}{| l | r | r | r | r |}
\hline
\textbf{System} & {\textbf{Blocks}} & {\textbf{Latency}} & {\textbf{StdDev}} &  {\textbf{Xput}} \\
\hline
Timestamping & 9638 & 0.2502s & 0.0772s &  229/s \\
\hline
Blockchain   & 120 & 2.72s & 0.31s &  0.37/s \\
\hline
\end{tabular}

\

By eschewing traditional blockchain design, our timestamping
 application has significantly better throughput and latency.
What's more, it requires only that the timestamping servers can
 reference the blocks being timestamped.
They do not have to interfere at all with other applications.
For instance, \textbf{our timestamping service can run on the storage
 nodes of our object model blockchains, providing each block with
 timestamps, with no measurable slowdown to the chains}.
Because our timestamping servers issue timestamps only when they have
 enough new blocks, the slower pace of the chains did slow down the
 rate of timestamping.

\section{Implementation}
\label{sec:implementation}
We have implemented a proof-of-concept version of Charlotte, and both
 timestamping~\p{\autoref{sec:timestamping}} and object
 model~\p{\autoref{sec:object}} applications.
We implemented a standard block format and an API for Fern and Wilbur
 servers in Apache Thrift, a datatype and network API
 language~\cite{Slee2007,thrift}.
Thrift supports encoding, decoding, and network communication in a
 wide variety of programming langauges~\cite{thrift}.
The specification comprises 473 lines of Thrift code, excluding
 comments, whitespace, and closing braces.

We implemented Wilbur and Fern servers in
 Haskell~\cite{haskell2010}.
Excluding comments, whitespace, and import statements, 
 our Wilbur implementation is 597 lines of code.
Our Fern implementation is 
 1057 lines of Haskell, including machinery specific to dealing with
 chains as defined in~\autoref{sec:chains}.
Each principal is identified by a 1024-bit RSA X509
 certificate~\cite{x509,rsa}.

Our Fern servers run a novel consensus algorithm based on byzantine
 Paxos~\cite{paxos-made-simple,byzantizing-paxos}.
It has two features useful for Charlotte, which we have not found
 elsewhere:
\ICS{This may be made simpler once we have gone back and specified
      what the integrity / consensus specification requirements are
      for Charlotte.}
\begin{itemize}
\item It produces a \textit{proof of consensus}~\p{\autoref{sec:consensus}}.
\item It has a notion of a consensus specification
       (an integrity label), as well as meets and
       joins~\p{\autoref{sec:joins}} on those specifications, allowing
       it to enforce chain properties with multi-chain
       blocks~\p{\autoref{sec:chain}}.
      Intuitively, the quorums for the meet consensus specification
       each feature one quorum from each of the labels'
       specifications.
\end{itemize}
The consensus implementation itself, excluding comments, whitespace,
 and imports, is 
 1796 lines of Haskell.

  \subsection{Experimental Setup}
  \label{sec:experimental}
  We ran sample chains on a cluster managed by
 Eucalyptus~\cite{eucalyptus} (software meant to simulate Amazon's
 cloud computing infrastructure~\cite{aws}).
Each machine has two 8-core Intel Xeon E5-2690 2.9 GHz
 CPUs~\cite{xenonE52690}.
The servers are connected by 10 GBit Ethernet.
Each of our servers was on a separate VM\footnote{
  VM specifications were chosen to fit within the resources of our
   cluster.},
 shown in the following table:

\ 

\begin{tabular}{| l | l | c | r |}
\hline
\textbf{Role} & \textbf{VM Type}  & \textbf{Cores} & \textbf{RAM}  \\
\hline
Client  & m1.large  & 4 & 16GB  \\
\hline
Fern    & m1.large  & 4 & 16GB  \\
\hline
Wilbur  & m2.xlarge & 2 & 2GB \\
\hline
\end{tabular}

\ 

\paragraph{The experiment client}
 mints and submits blocks for all the chains.
Each block is begun as soon as attestations are available for 
 references to the previous blocks.
\paragraph{Fern servers}
 run a 4-participant consensus tolerating 1 byzantine failure for each
 chain.
\paragraph{Wilbur servers}
 each store blocks for a specific chain.
Each chain has 4 servers, and blocks must be stored on 3.

  \subsection{Performance}
  \label{sec:performance}
  We ran a series of experiments to demonstrate the viability of
 applications using the Object Model~\p{\autoref{sec:object}}. 
Each experiment begins with a root block for each chain.
We append a block to all chains together, 12 times in a row.
We repeated this experiment 10 times for 1, 2, and 3 chains.
The mean times to append each of the 120 blocks are in the following
 table:

\ 

\begin{tabular}{| r | r | r | r |}
\hline
\textbf{Chains} & \textbf{Time (s)} & \textbf{StdDev} & \textbf{StdErr} \\
\hline
1 & 2.72 & 0.31 & 0.03 \\
\hline
2 & 15.7 & 2.4 & 0.2 \\
\hline
3 & 196 & 82 & 8 \\
\hline
\end{tabular}

\ 

While these results are not fast, they are more than sufficient to
 demonstrate simple blockchains.
We believe much faster performance is possible with more optimized
 parsing and consensus.

  \subsection{Discussion}
  \label{sec:discussion}
  We believe much faster implementations of Charlotte are possible with
additional engineering effort.
Profiling reveals that the current prototype spends a
 significant portion of its CPU time marshaling and unmarshaling data.
Our prototype implementation uses the Thrift marshaling and RPC
 system~\cite{Slee2007,thrift}.
We created a microbenchmark to test the relative performance of the
 similar RPC system gRPC~\cite{grpc} using Protobufs~\cite{protobufs}.
For our message sizes, gRPC moves structured data
 20--30$\times$ faster;
even with no other changes, a reimplementation based on Protobufs
 should yield sub-second multichain transaction times.

Marshaling is a dominant factor in part because our consensus
algorithm leads to large message sizes.
 Fortunately, these message sizes could be brought down substantially
 by removing redundancy. For example,
to simplify construction of proofs of consensus, each message in our
 consensus system carries the original proposed
 value, which does not need to be retransmitted with every message.

Other small improvements, including selecting more compact
 cryptographic ids and signatures, and representing
 integrity labels less verbosely, could improve our implementation.
While Haskell yields concise, clear code,
a performance-oriented implementation might choose differently.
\ACM{I think we emphasize that the real figure of merit is the
number of rounds. Also, do we report that out clearly?}

\section{Evaluation}
\label{sec:evaluation}
\begin{figure}
  \centering
  {\small
\begin{tabular}{| l   r   l | l | l |}
\hline
& & & \textbf{Unaltered} & \textbf{2 Accounts} \\
\hline
\multirow{3}{*}{serialized}   & \multicolumn{2}{r|}{longest chain}      &
\bitcoinnumtransactions & \bitcoinTwoTXTotal
\\\cline{2-5}
                              & \multirow{2}{*}{speed}      & \bitcoin  &
\bitcointotaltime & \bitcoinTwoTXTotalTime
\\\cline{3-5}
                              &                             & Prototype &
N/A & \bitcoinTwoTXExperimentTime
\\\hline
\multirow{3}{*}{parallelized} & \multicolumn{2}{r|}{longest chain}      &
\bitcoinlengthlongest & \bitcoinlengthlongestweighted
\\\cline{2-5}
                              & \multirow{2}{*}{speed}      & \bitcoin  &
\bitcoincharlottetotaltime & \bitcoincharlotteweightedtotaltime
\\\cline{3-5}
                              &                             & Prototype &
N/A & \bitcoincharlotteweightedexperimenttime
\\\hline
\end{tabular}
}

  \vspace{6mm}
  \caption{Payment data from \bitcoin's first {\bitcoinnumblocks} blocks \\
           (We were unable to evaluate our prototype on transactions of arbitrarily many accounts.)}
  \label{fig:evaluation}
\end{figure}

We evaluate the potential efficiency savings of Charlotte on a payment
 network by analyzing first {\bitcoinnumblocks} of the real {\bitcoin}
 transaction history~\cite{blockchain-info}.
Our measurements are based on our prototype implementation,
 which has been tested running chains in our object
 model~\p{\autoref{sec:object}}.

Our Charlotte banking model is based on our object model: each bank
 account is a chain.
This presents some difficulty in direct comparison to {\bitcoin}, as
 {\bitcoin} doesn't keep track of money in terms of accounts.
Instead, each transaction divides all its money into a number of
 outputs, each of which specify the conditions under which they can be
 spent
 (such as need a signature matching a specific public key\ldots).
These are called Unspent Transaction Outputs, or UTXOs.
Each transaction specifies a set of input UTXOs as well, from whence
 it gets the money, and provides for each a proof that it is
 authorized to spend the money.
Each UTXO is completely drained when it is spent, and cannot be
 re-used.
Thus the transactions in {\bitcoin} form a graph, with transactions as
 vertices, and UTXOs as directed edges~\cite{bitcoin}.

In Charlotte, transactions need only be approved by the servers
 specified by the participants.
These are likely to be a handful of bank servers (enough to tolerate a
 small number of failures), executing a consensus protocol.
Paxos-based protocols, like our implementation, have a 3-message
 latency for commit, absent conflicting commits.
By contrast, {\bitcoin} requires at least 6 instances of gossiping
 blocks through the entire network to commit securely, and each
 instance requires many message sends~\cite{bitcoin}.
With only a few servers running consensus for each transaction,
 Charlotte would also consume several orders of magnitude less energy
 than a proof-of-work based blockchain~\cite{bitcoinenergy}.

\ 

\begin{tabular}{| l | r | r |}
\hline
\textbf{communication / TX} & \textbf{energy / TX} & \textbf{Machines}\\\hline
3 message sends &   $\sim 3000000$ kJ  & $\sim 50,000$ \\\hline
6 gossip instances &     $\sim 30$ kJ  & $\sim 5$ / bank \\\hline
\end{tabular}

\ 

In our Charlotte banking model, transfers between two accounts are
 simply a block on both chains.
Therefore, so far as two sets of financial transactions don't
 interact, they can operate entirely in parallel.
The time it takes to make all the payments in the dataset is therefore
 the maximum time it takes to make the payments corresponding to any
 single path through the payment graph.
This path of payments, each referencing the previous, corresponds to a
 chain of blocks.
If all transactions are between exactly two accounts, and it always
 takes the same amount of time to append a block to two chains, then
 the time it takes to make the longest chain, and thus the time it
 takes to make all the payments in the dataset, is proportional to the
 length of the longest path.
\ICS{This sentence is now really long.
     If anyone has a good idea on how to break it up, please do.}

The speed analysis of Charlotte's parallelized versus {\bitcoin}'s
 serialized approaches is in~\autoref{fig:evaluation}.
In principle, Charlotte only needs time to order
 {\bitcoincharlottepercenttime} of the transactions {\bitcoin} orders.
If they commit transactions at the same rate, our Charlotte banking
 design would take {\bitcoincharlottepercenttime} of the time
 {\bitcoin} would to complete a group of payments.

However, that assumes each transaction takes the same amount of
 time to commit.
In {\bitcoin}, it improves anonymity and performance to
combine many small transfers of
 money into big ones, with many inputs and many outputs.
However, in Charlotte, more accounts means more servers trying to
 achieve consensus.
In most consensus implementations, more participants make consensus
 substantially slower~\cite{LiuLKA2016,Miller2016}; our implementation
 is no exception.

In the American financial system, all
 monetary transfers are from one account to another.
Blocks are effectively limited to two chains each.
We can simulate this limitation by refactoring each \bitcoin transaction
 as a DAG of transactions with depth logarithmic in the number
 of participants (see \autoref{sec:twoaccounts}).

The speed analysis for this construction is
 in~\autoref{fig:evaluation}, in the ``2 Accounts'' column.
If Charlotte committed transactions at the same speed {\bitcoin} does,
 it would take only {\bitcoincharlotteweightedpercenttime} of the time
 that {\bitcoin} takes to commit a group of transactions, even after
 refactoring the transactions.

However, Charlotte's transactions need only satisfy the requirements
 of the bank accounts involved, allowing them to specify faster
 consensus mechanisms.
Even with our unoptimized proof-of-concept implementation, with each
 chain running a 4-participant byzantine-tolerant consensus among
 its Fern servers, we append a block to two chains in
 $15.7\pm0.2$ seconds.
What took the {\bitcoin} system \bitcointotaltime{} would thus take
 Charlotte \bitcoincharlotteweightedexperimenttime{}, a mere
 {\bitcoincharlotteweightedexperimenttimepercent} of {\bitcoin}'s time.
We believe that faster consensus implementations are
 possible, allowing Charlotte to become even more
 efficient.

\ifextensions
\section{Extensions}
\ACM{Lots of nice ideas here.
    I suspect we should just save them for a future paper and use the
     space here to explain all the existing great ideas more clearly
     and convincingly.
}\ICSreply{You're probably right.
           I suppose I just thought some of these were cool.
           I've left this section in for the moment, but if and when
            we want it gone, just comment out this part of body.tex.}
\label{sec:extensions}
Charlotte is designed to be extensible.
New servers, applications, and features can be added without harming
 existing ones.
We have considered several features which might be added to further
 enhance Charlotte's advantages.
  \subsection{Confidentiality}
  \label{sec:confidentiality}
  In principle, we can add confidentiality policies to a block's label,
 stating that references to a block must be encrypted such-and-such a
 way.
If anyone has a signed reference (e.g., part of a block someone's
 attested to), which includes the label and thus the confidentiality
 policy, not encrypted in such a way, they can prove someone leaked.

The contents of a block can be encrypted, providing a certain measure
 of confidentiality.
Naturally, if the label is encrypted, it may be difficult for Fern
 servers to attest to anything about the block, or for anyone to
 compare such attestations to the label, without the ability to
 decrypt it.
There may be cryptographic techniques for side-stepping this
 difficulty~\cite{Kosba2016}.

While it may seem obvious that block contents can be encrypted, it is
 noteworthy that \textit{the existence of a block itself} can be kept
 confidential.
Labels could be extended with other policies, including a
 confidentiality policy detailing the encryption requirements of any
 reference to the block.
A reference that is not encrypted to the standards set out in the
 label (which is in the reference itself) is a proof of leaking
 information.
Anyone who signs anything featuring such a reference has provably
 violated the confidentiality policy.
While it is impossible to prevent those with knowledge of a block from
 leaking it out-of-band, it is at least possible to disincentivize
 improper references within Charlotte itself.

For instance, a block might specify: ``no reference to this block
 should be decryptable without a private key matching this public
 key.''
Anyone signing a reference which can be shown to be decryptable,
 without such a private key, can prove that whoever published the
 reference leaked.

The default confidentiality policy, applied to all blocks without a
 confidentiality policy in their label, would be that references can
 be public and unencrypted.

  \subsection{Block Deletion}
  \label{sec:deletion}
  Wilbur servers might add more particulars to their availability
 attestations.
Instead of a signed promise, or proof of retrievability promising to
 provide the block forever, they might specify the circumstances under
 which they may respond to requests with something other than the
 block.
For instance, a Wilbur server might say that it promises to provide
 this block, which claims slot 7 on a chain, unless it can respond to
 requests for this block with attestations from the chain's Fern
 servers for a different block claiming slot 7.
Since references to a block include Wilbur attestations, any Wilbur
 server could respond to a request to fill a reference with any
 information it knows that satisfies the deletion policy, instead of
 the block.
This could substantially cut down on storage space.

We can add deletion policies to a block's availability label as well.
These would specify what terms the block requires from Wilbur servers
 (beyond a simple ``store it forever'').

The default deletion policy, applied to all blocks without such a
 thing in their label, would be that Wilbur servers must vow to store
 a block permanently.

  \subsection{Compacting Attestations}
  \label{sec:compacting}
  Advanced cryptographic techniques may be able to compact several
 attestations within the same block reference into a smaller number of
 bits~\cite{Bellare2006}.
Furthermore, recursive attestations would naively require that the
 entire chain from the proof to its ancestor block be presented, but
 it may be possible to compact these as well~\cite{Kiayias2016}.
Recall that recursive attestations, which include proof of work, are
 attestations for a block which implicitly attest to the block's
 ancestors~\p{\autoref{sec:chain}}.

\fi 

\section{Related Work}
\label{sec:related}
Although the original {\bitcoin} protocol offers limited support for
 smart contracts~\cite{bitcoin, ivy-bitcoin}, far more developers have
 flocked to {\ethereum}'s general-purpose platform for communicating
 applications~\cite{ethereum, Dhillon2017}\ICS{
   Dhillon2017 is here because I needed someone to cite for the fact
    that {\ethereum} is popular, and their whitepaper isn't just a
    pipe dream.
   If there's a better source, use that.}.
Unlike {\bitcoin} and {\ethereum}, however, Charlotte does not
 constrain all blocks to be in one chain, leaving applications free of
 unnecessary serialization, and free to use less expensive consensus
 mechanisms.

Charlotte is not the first blockchain project to propose separating
 availability and integrity servers.
\blockchainName{StorJ}~\cite{storj} and
 \blockchainName{Filecoin}~\cite{filecoin} separate the notion of the
 blockchain itself entirely from the data being stored.
\blockchainName{BigchainDB}~\cite{bigchaindb}, however, separates its
 byzantine-tolerant transaction ordering service from the underlying
 storage for those transactions, for which it uses
 MongoDB~\cite{membrey10}.
Charlotte takes this one step further, allowing blocks of arbitrary
 size (even for file storage) to be stored on Wilbur servers, without
 requiring that Fern servers see whole blocks at all.

  \paragraph{Unchained blocks:}
  \label{sec:unchained}
  Several current projects involve adapting blockchains into non-chain
 structures.
\blockchainName{spectre}~\cite{Sompolinsky2016},
 \blockchainName{iota}~\cite{iota,Popov2017}, and
 \blockchainName{nano} (a.k.a. \blockchainName{raiblocks})~\cite{nano}
 are each DAGs of blocks supporting a cryptocurrency,
 but are not general application frameworks.

Several projects endeavor to provide the illusion of a single chain,
 while dividing the work amongst several shards.
\blockchainName{elastico} is a sharded proof-of-work chain which
 divides miners into semi-independent shards to be unified
 each ``epoch''~\cite{Luu2016}.
{\ethereum} is considering ways to shard
 its workload to handle more throughput~\cite{ethereum-sharding}.
\blockchainName{zilliqa} is a sharded blockchain in
 testing~\cite{zilliqa}.
It is in many ways like {\ethereum}, although
it still requires full nodes storing all global state,
and each transaction must be parallelizable across affected shards.
All of these projects rely on a global, proof-of-work based miner
 membership scheme, and attempt to distribute work automatically.
Charlotte allows applications to choose more scalable storage and
 consensus schemes, and provides a principled framework in which
 different chains can interoperate.

  \paragraph{Integrity meets:}
  \label{sec:meets}
  To append a block to multiple Charlotte chains, the new block requires
 integrity attestations satisfying both chains.
While it is possible to seek approval for each separately, this runs
 the risk that one chain will approve the block, while the other will
 approve a conflicting block, leaving the first chain stuck.
Charlotte solves this problem precisely when the integrity policies of
 both chains have a \textit{meet}, or ``least upper bound,'' which is
 a new integrity policy satisfying both~\p{\autoref{sec:joins}}.
This concept is akin to the integrity necessary to write multiple
 information-flow labeled objects~\cite{jif}\ICS{
   What all should I cite here?}.
The consensus we use in our proof-of-concept implementation has meets,
 and so a single consensus for both chains is run whenever we append a
 block to two chains, and the block is approved for both, or neither.

Other systems face the challenge of merging data at different
 integrity levels, with various solutions.
Projects building sharded blockchains discuss strategies for inter-shard
 transactions, many of which can be used in
 Charlotte~\cite{ethereum, zilliqa}.
To our knowledge, none have implemented them.
Some, such as \blockchainName{Elastico}~\cite{Luu2016}, avoid the
 problem by trusting shards equally, and focusing solely on a
 crypto-currency designed so inter-shard transactions are unnecessary.
The interledger protocol~\cite{interledger} uses a kind of 2--phase
 commit~\cite{2pc} to atomically commit a transaction to multiple
 chains.
The advantage of this technique is that it works for almost any
 consensus mechanism on each chain.
The disadvantage is that each chain must have some way to interpret
 the special transactions which the protocol uses, and it takes at
 least two rounds of consensus on each chain to commit.
For these reasons, such a protocol could be used on Charlotte chains,
 but sometimes more convenient integrity meets may be available.

\section{Conclusion}
\label{sec:conclusion}
Charlotte is a decentralized \blockweb, to which anyone can contribute
 blocks, servers, and applications.
Since each block's approval process only has to satisfy the parties
 involved, no expensive global consensus mechanism is required.
Combined with avoiding unnecessary serialization, this
 means Charlotte applications can run far faster and more efficiently
 than their traditional blockchain counterparts, while retaining many
 of the same guarantees.
Since the duties of availability and integrity are
 separated, more specialized services can more efficiently cater to
 each.

The extensible format of Charlotte blocks allows each application to
 use whichever consensus mechanism or availability attestation it wants.
This makes it an ideal framework in which to build new blockchains,
 both experimental and industrial.
We plan to release the source code of our proof-of-concept servers,
which we hope will bootstrap the adoption of the \blockweb.

\ifacknowledgments


\fi

\appendix

\section{Additional Applications}
\label{sec:additional}
  \subsection{Version Control}
  \label{sec:version}
  Version control systems such as Git~\cite{Chacon2014} already
 maintain \textit{commits}, or versions of the codebase, as a DAG.
Each commit references whatever prior commits it was based on.
In Git, a commit with multiple references is called a ``merge.''

These systems have no inherent limit to who can mint blocks, or with
 what predecessors.
Two things require agreement:
\begin{itemize}
\item Pushing a commit to a server, in which case the server must
       choose to accept it, in essence promising to provide that
       commit to others, when requested.
\item Designating a commit as part of a \textit{branch}, mutable names
       each machine assigns to one commit at a time.
\end{itemize}
These two correspond exactly to the attestations of Wilbur and Fern
 servers.
The former is simply block storage, while in the latter, branches
 correspond to attestations~\p{\autoref{sec:attestations}}.
When a machine attests that one commit is its latest choice for a
 branch, that commit must refer to its previous choice.
Each time it assigns a new commit to a branch, it attests never to
 assign a different commit to that branch, unless this one is in its
 ancestry.
In fact, while the duties of the two are usually shared by the same
 machines (e.g. github~\cite{github}), projects like
 Git~LFS~\cite{gitlfs} separate along exactly the same lines as
 Wilbur and Fern.
The advantage of making a version control system compatible with a
 larger Charlotte {\blockweb} would be composability.
Version control users could use general-purpose Wilbur storage servers
 to store code, and other applications referencing commits or branches
 need no version-control-specific mechanism to do so.

\ICS{Used to include the following, not sure it's still useful:
Furthermore, attestations can take on a new meaning in the context of
 version control.
A Fern server representing, say, a code reviewer, could attest to only
 those commits with code that passes muster.
In essence, the Fern server maintains a data structure which is the
 set of commits that pass review.
Branches, being chains, might require such an attestation in their
 label, effectively building code review into the version control system.
}

  \subsection{Medical Records}
  \label{sec:medical}
  Several recent blockchain efforts are focused on medical
 records~\cite{Marr2017,Azaria2016,Mettler2016,Liu2016}.
Here again, there is no need for total serialization of all records.
One simple model might be to make each record an object, maintained
 by Fern and Wilbur servers of the patient's choice.
Doctors authorized by the patient or relevant authorities contribute
 transactions to the object.
This would mimic the existing USA medical record system, in which the
 authoritative record for a patient is kept by their ``primary
 physician''~\cite{hippa}.
The primary physician's record-keeping duties would simply be made
 electronic.

Like our version control example~\p{\autoref{sec:version}},
 supervising doctors or hospitals could attest to certain records,
 indicating that they were kept properly.
This allows automatic enforcement of some policies, such as allowing
 some nurses to prescribe medication if a doctor signs off on it.

\section{Bitcoin Transactions in Two Accounts or Less}
\label{sec:twoaccounts}
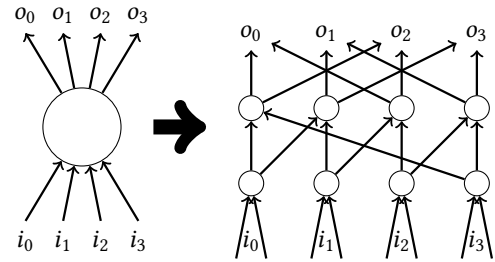
\begin{figure}
  \centering
  \begin{tikzpicture}
\draw (.75,1.75) node[draw, circle, text width=5mm, text height = 5mm] (central) {};

\draw[->, line width=0.3mm] node[anchor=north] at (0,.5) {$i_0$} (0,.5) -- (central);
\draw[->, line width=0.3mm] node[anchor=north] at (.5,.5) {$i_1$} (.5,.5) -- (central);
\draw[->, line width=0.3mm] node[anchor=north] at (1,.5) {$i_2$} (1,.5) -- (central);
\draw[->, line width=0.3mm] node[anchor=north] at (1.5,.5) {$i_3$} (1.5,.5) -- (central);

\draw[->, line width=0.3mm] (central) -- (0,3) node[anchor=south] at (0,3) {$o_0$};
\draw[->, line width=0.3mm] (central) -- (.5,3) node[anchor=south] at (.5,3) {$o_1$};
\draw[->, line width=0.3mm] (central) -- (1,3) node[anchor=south] at (1,3) {$o_2$};
\draw[->, line width=0.3mm] (central) -- (1.5,3) node[anchor=south] at (1.5,3) {$o_3$};

\draw[->, line width=2mm] (1.7,1.75) -- (2.4, 1.75);

\draw (3,1) node[draw, circle] (c00) {};
\draw (3,2) node[draw, circle] (c01) {};
\draw (3,3) node[] (c02) {$o_0$};

\draw (4,1) node[draw, circle] (c10) {};
\draw (4,2) node[draw, circle] (c11) {};
\draw (4,3) node[] (c12) {$o_1$};

\draw (5,1) node[draw, circle] (c20) {};
\draw (5,2) node[draw, circle] (c21) {};
\draw (5,3) node[] (c22) {$o_2$};

\draw (6,1) node[draw, circle] (c30) {};
\draw (6,2) node[draw, circle] (c31) {};
\draw (6,3) node[] (c32) {$o_3$};

\draw[->, line width=0.3mm] (2.8,0) -- (c00);
\draw[->, line width=0.3mm] node[anchor=north] at (3,.5) {$i_0$};
\draw[->, line width=0.3mm] (3.2,0) -- (c00);

\draw[->, line width=0.3mm] (3.8,0) -- (c10);
\draw[->, line width=0.3mm] node[anchor=north] at (4,.5) {$i_1$};
\draw[->, line width=0.3mm] (4.2,0) -- (c10);

\draw[->, line width=0.3mm] (4.8,0) -- (c20);
\draw[->, line width=0.3mm] node[anchor=north] at (5,.5) {$i_2$};
\draw[->, line width=0.3mm] (5.2,0) -- (c20);

\draw[->, line width=0.3mm] (5.8,0) -- (c30);
\draw[->, line width=0.3mm] node[anchor=north] at (6,.5) {$i_3$};
\draw[->, line width=0.3mm] (6.2,0) -- (c30);


\draw[->, line width=0.3mm] (c00) -- (c01);
\draw[->, line width=0.3mm] (c00) -- (c11);

\draw[->, line width=0.3mm] (c10) -- (c11);
\draw[->, line width=0.3mm] (c10) -- (c21);

\draw[->, line width=0.3mm] (c20) -- (c21);
\draw[->, line width=0.3mm] (c20) -- (c31);

\draw[->, line width=0.3mm] (c30) -- (c31);
\draw[->, line width=0.3mm] (c30) -- (c01);

\draw[->, line width=0.3mm] (c01) -- (c02);
\draw[->, line width=0.3mm] (c01) -- (c22);

\draw[->, line width=0.3mm] (c11) -- (c12);
\draw[->, line width=0.3mm] (c11) -- (c32);

\draw[->, line width=0.3mm] (c21) -- (c22);
\draw[->, line width=0.3mm] (c21) -- (c02);

\draw[->, line width=0.3mm] (c31) -- (c32);
\draw[->, line width=0.3mm] (c31) -- (c12);
\end{tikzpicture}
  \caption{Converting 4 inputs and 4 outputs to a graph of 2-account transactions.}
  \label{fig:twoaccounts}
\end{figure}

In {\bitcoin}, it is advantageous to combine many small transfers of
 money into big ones, with many inputs and many outputs.
 \ACM{Not sure this appendix is actually worth the space it takes up}
 \ICSreply{From a page limit standpoint, appendices don't take up
            space.}
This improves anonymity and performance.
In the real financial system of the USA, however, all monetary
 transfers are from one account to another.
They are all exactly two chain transactions.

We can simulate this limitation by refactoring each {\bitcoin} UTXO as
 2 UTXOs, and each {\bitcoin} transaction as a DAG of transactions with
 depth:
\[
\left\lceil log_2\p{max\p{\textrm{number of inputs}, \textrm{ number of outputs}}} \right\rceil
\]

To do this, we create 
\[
  n := 2^d
\]
 chains, each of which is
\[
  d := \left\lceil log_2\p{max\p{\textrm{number of inputs}, \textrm{ number of outputs}}} \right\rceil
\]
 long.
We call these chains $C^0$ through $C^n$.
Original input UTXO $i$ corresponds to both inputs to the first
 transaction of chain $i$.
Original output UTXO $j$ corresponds to one output of each of the last
 transactions from chains $j$ and $\p{j + 2^{d-1}}\textrm{mod }n$.
For $0 \leq k < \p{d-1}$, the outputs of the $k^{th}$ transaction in chain
 $i$, called $C_k^i$, go to $C_{k+1}^i$, and:
\[
  C_{k+1}^{\p{i + 2^j} \textrm{mod}\ n}
\]
The outputs of $C_d^i$ go to the UTXOs corresponding with output $i$,
 and output $\p{i + 2^{d-1}}\textrm{mod }n$.
Each transaction divides its output values proportionately to the sums
 of the final output values reachable from each of the transaction's
 outputs.
\autoref{fig:twoaccounts} is an example transformation from a 4-input,
 4-output transaction to a DAG of depth 2 using all 2-input, 2-output
 transactions.

\bibliography{bibtex/pm-master}

\end{document}